\documentclass[12pt,preprint]{aastex}

\def\ms{m~s$^{-1}$}
\def\ks{km~s$^{-1}$}
\def\fe{$\rm [Fe/H]$}
\def\vsini{$V_{rot}\sin{i}$}
\def\msini{$M_P\sin{i}$}
\def\asini{$a\sin{i}$}
\def\logg{$\log{g}$}
\def\msun{$M_{\odot}$}
\def\mjup{$M_{\rm Jup}$}
\def\rsun{$R_{\odot}$}
\def\chisq{$\sqrt{\chi^2_\nu}$}
\def\plmn{~$\pm$~}

\def\nksig{4.47}

\def\teffa{6028}
\def\fea{0.257}
\def\ga{4.36}
\def\va{4.2}
\def\rsa{1.22}

\def\teffc{6037}
\def\fec{0.343}
\def\gc{4.38}
\def\vc{3.5}
\def\rsc{1.70}

\def\teffd{5995}
\def\fed{0.366}
\def\gd{4.21}
\def\vd{3.2}
\def\rsd{1.20}

\def\pa{2.1375}
\def\pea{0.0002}
\def\tpa{2453694.8}
\def\tpea{0.3}
\def\ea{0.008}
\def\eea{0.004}
\def\ka{207.7}
\def\kea{0.8}
\def\oma{251}
\def\omea{40}
\def\ma{1.50}
\def\rmsa{3.20}
\def\chia{0.81}
\def\nobsa{26}
\def\aa{0.039}
\def\msa{1.21}
\def\vmaga{8.73}
\def\mva{3.93}
\def\bva{0.664}
\def\spa{F8V}
\def\da{91}
\def\shka{0.16}
\def\rhka{-5.03}
\def\pra{40.1}

\def\pc{26.73}
\def\pec{0.02}
\def\tpc{2453193.9}
\def\tpec{3.}
\def\ec{0.05}
\def\eec{0.03}
\def\kc{40.2}
\def\kec{2.}
\def\omc{6.}
\def\omec{200}
\def\mc{0.71}
\def\rmsc{4.07}
\def\chic{0.85}
\def\nobsc{24}
\def\ac{0.233}
\def\msc{1.33}
\def\vmagc{8.23}
\def\mvc{3.36}
\def\bvc{0.639}
\def\spc{G2IV}
\def\dc{94}
\def\shkc{0.13}
\def\rhkc{-5.15}
\def\prc{27.4}

\def\pd{18.179}
\def\ped{0.007}
\def\tpd{2453017.6}
\def\tped{0.3}
\def\ed{0.48}
\def\eed{0.05}
\def\kd{25.2}
\def\ked{2.}
\def\omd{155.8}
\def\omed{8.}
\def\md{0.33}
\def\rmsd{3.60}
\def\chid{0.77}
\def\nobsd{25}
\def\ad{0.168}
\def\msd{1.24}
\def\vmagd{8.05}
\def\mvd{3.36}
\def\bvd{0.641}
\def\spd{G3V}
\def\dd{86}
\def\shkd{0.10}
\def\rhkd{-5.60}
\def\prd{55.5}

\def\pam{2.1378}
\def\peam{0.0002}
\def\tpam{2453694.4}
\def\tpeam{0.4}
\def\eam{0.007}
\def\eeam{0.005}
\def\kam{207.8}
\def\keam{1.0}
\def\omam{192}
\def\omeam{70}
\def\mam{1.50}
\def\rmsam{4.38}
\def\chiam{0.87}
\def\nobsam{26}
\def\aam{0.039}

\begin{document}
\title{The N2K Consortium VI: Doppler Shifts Without Templates
  and Three New Short--Period Planets~$^1$}

\author{ John Asher Johnson\altaffilmark{2}, 
  Geoffrey W. Marcy\altaffilmark{2},
  Debra	A. Fischer\altaffilmark{3}, 
  Gregory Laughlin\altaffilmark{4}, 
  R. Paul Butler\altaffilmark{5},
  Gregory W. Henry\altaffilmark{6},
  Jeff A. Valenti\altaffilmark{7},
  Eric B. Ford\altaffilmark{2}, 
  Steven S. Vogt\altaffilmark{4},
  Jason T. Wright\altaffilmark{2} 
 }
\email{  johnjohn@astron.berkeley.edu }

\altaffiltext{1}{Based on observations obtained at the W. M. Keck
Observatory, which is operated jointly by the University of California and
the California Institute of Technology}
\altaffiltext{2}{Department of Astronomy, University of California,
Mail Code 3411, Berkeley, CA 94720}
\altaffiltext{3}{Department of Physics \& Astronomy, San Francisco
  State University, San Francisco, CA 94132}
\altaffiltext{4}{UCO/Lick Observatory, University of California at
  Santa Cruz, Santa Cruz, CA 95064}
\altaffiltext{5}{Department of Terrestrial Magnetism, Carnegie
  Institution of Washington DC, 5241 Broad Branch Rd. NW, Washington DC,
 20015-1305}
\altaffiltext{6}{Center of Excellence in Information Systems, Tennessee
  State University, 3500 John A. Merritt Blvd., Box 9501, Nashville, TN 37209}
\altaffiltext{7}{Space Telescope Science Institute, 3700 San Martin Dr., Baltimore, MD 21218}

\begin{abstract}
We present a modification to the iodine cell
Doppler technique that eliminates the need for 
an observed stellar template spectrum. For a given target
star, we iterate toward a synthetic template spectrum 
beginning with an existing spectrum of a similar star. We 
then perturb the shape of this first--guess template to match 
the program observation of the target star taken through an 
iodine cell. The elimination of a separate template observation 
saves valuable telescope time, a feature that is ideally 
suited for the quick--look strategy employed by the 
``Next 2000 Stars'' (N2K) planet search program.  Tests 
using Keck/HIRES spectra indicate that synthetic templates 
yield a short--term precision of 3~\ms\ and a long--term, 
run--to--run precision of 5~\ms. We used this new Doppler
technique to discover three new planets: a \ma~\mjup\ planet in a \pa~d
orbit around HD~86081; a \mc~\mjup\ planet in circular, \pc~d
orbit around HD~224693; and a Saturn--mass planet in an \pd~d orbit
around HD~33283. The remarkably short period of HD~86081b bridges 
the gap between the extremely short--period planets detected in the
OGLE survey and the 16 Doppler--detected hot jupiters ($P < 15$~d),
which have an orbital period distribution that piles
up at about three days. 
We have acquired photometric observations of two of the planetary host stars 
with the automated photometric telescopes at Fairborn
Observatory.  HD~86081 and HD~224693 both lack detectable 
brightness variability on their radial velocity periods, supporting 
planetary-reflex motion as the cause of the radial velocity variability.  
HD~86081 shows no evidence of planetary transits in spite of a 17.6\% transit 
probability.  We have too few photometric observations to detect or rule out 
transits for HD~224693.
\end{abstract}


\section{Introduction}
Of the 171 planets that have been found
orbiting stars within 200~pc of the Sun\footnote{References to
  published papers and updates on orbital parameters can be found at
  http://exoplanets.org}, the vast majority of have been discovered
by measuring Doppler reflex motion of their parent stars
\citep{butler06}. Effectively detecting 
the relatively small Doppler shifts caused by planet-sized bodies
requires velocity precision of order 1~\ms. This corresponds
to measuring shifts in spectral lines to within $10^{-3}$ of a resolution
element for modern echelle spectrometers such as Keck HIRES
\citep{vogt94}. Such a feat requires intimate knowledge of wavelength
scale of the spectrometer detector. However, the wavelength scale is
not constant with time. The instrument is subject to changes on time
scales as short as hours or even minutes due to effects such as telescope
flexure and thermal expansion. It is therefore necessary to measure
the wavelength scale for each observation using a reference
spectrum obtained simultaneously with the stellar spectrum.    

One of the most successful and widely--used precision velocity methods
is the iodine cell technique \citep[][hereafter B96]{butler96}. The B96
technique uses an iodine cell placed in front of the
entrance slit of an echelle spectrometer, which imprints a dense set
of narrow molecular lines on each stellar spectrum. Velocity
information is extracted by fitting a model consisting of a stellar
template spectrum multiplied by an iodine reference spectrum and
convolved with an instrumental profile, or PSF. The shape of the PSF,
wavelength scale of the spectrometer, and the Doppler shift of the
stellar template are left as free parameters in the fit. 

The stellar template spectrum is an observation of the target star
obtained without the iodine cell, with the instrumental profile
deconvolved. Because numerical deconvolution algorithms are highly 
susceptible to noise amplification \citep[e.g.][]{starck02},
template observations must have high signal-to-noise (S/R~$ \gtrsim
500$) and high spectral resolution ($\lambda/\Delta\lambda \gtrsim 80,000$).
Both of these requirements result in longer exposure times compared to
program observations made through the iodine cell.

One observing program for which this telescope time overhead is 
prohibitive is the Next 2000 Stars (N2K) planet search
\citep[][hereafter Paper I]{fischer05a}. N2K is an international
consortium designed to search for short--period planets ($P<15$~d,
commonly known as ``hot jupiters'') orbiting
metal--rich stars residing within 110 parsecs of the Sun. As telescope
time is allocated at each of three major telescopes---Keck, Magellan and 
Subaru---each target star is monitored for three consecutive nights. Those
stars exhibiting radial velocity (RV) variations between 20 and 250~\ms~are
flagged for follow-up spectroscopic observations to search for a full
orbital solution. Planet candidates also receive photometric
monitoring with the automatic photometric 
telescopes (APT) at Fairborn Observatory \citep{henry99} to search for
transit events. To date, this "quick--look" observing strategy has
resulted in the detection of four short--period planets
\citep[Paper I;][]{fischer06}, one of which, HD 149026, transits its
host star \citep{sato05}. 

Most N2K target stars receive only three RV observations,
which means that template observations represent more than 25\% of the
telescope time required for the program. Since template
observations contain no inherent velocity information, it would be
highly advantageous to bypass them by measuring Doppler shifts
with respect to synthetic reference spectra. Using this motivation,
we have developed a modification to the B96 velocity method whereby we
derive the shape of the intrinsic stellar spectrum directly from 
normal iodine cell observations. This synthetic, or ``morphed,'' template is
then used in place of a deconvolved template in the B96
modeling process.  

We present an overview of the forward modeling
procedure used to produce morphed templates in
\S~\ref{method}. In \S~\ref{testing} we show the results of several
tests of both the 
short-- and long--term precision attainable from morphed templates
using Keck/HIRES observations of a set of velocity standard stars. In
\S~\ref{application} we detail how this new method is currently being used
in the N2K planet search. We also report the detection of three new
short--period planets, and describe the results of our photometric
monitoring of the host stars for transit events. Finally, in
\S~\ref{discussion} we summarize our results, propose other potential
applications for spectral morphing and discuss the new planet
discoveries with respect to the growing statistical ensemble of
short--period exoplanets. 

\section{Spectral Morphing}
\label{method}

The B96 precision velocity technique employs a
temperature--controlled pyrex cell 
containing gaseous molecular iodine placed at the entrance slit of a
spectrometer. The iodine lines imprinted on the stellar spectrum serve 
as a robust wavelength fiducial against which the Doppler shift of the
stellar lines is measured. The Doppler information from each iodine
observation is extracted using a forward modeling algorithm described
in detail by B96 and summarized below.

\begin{figure}[t!]
\epsscale{0.8}
\plotone{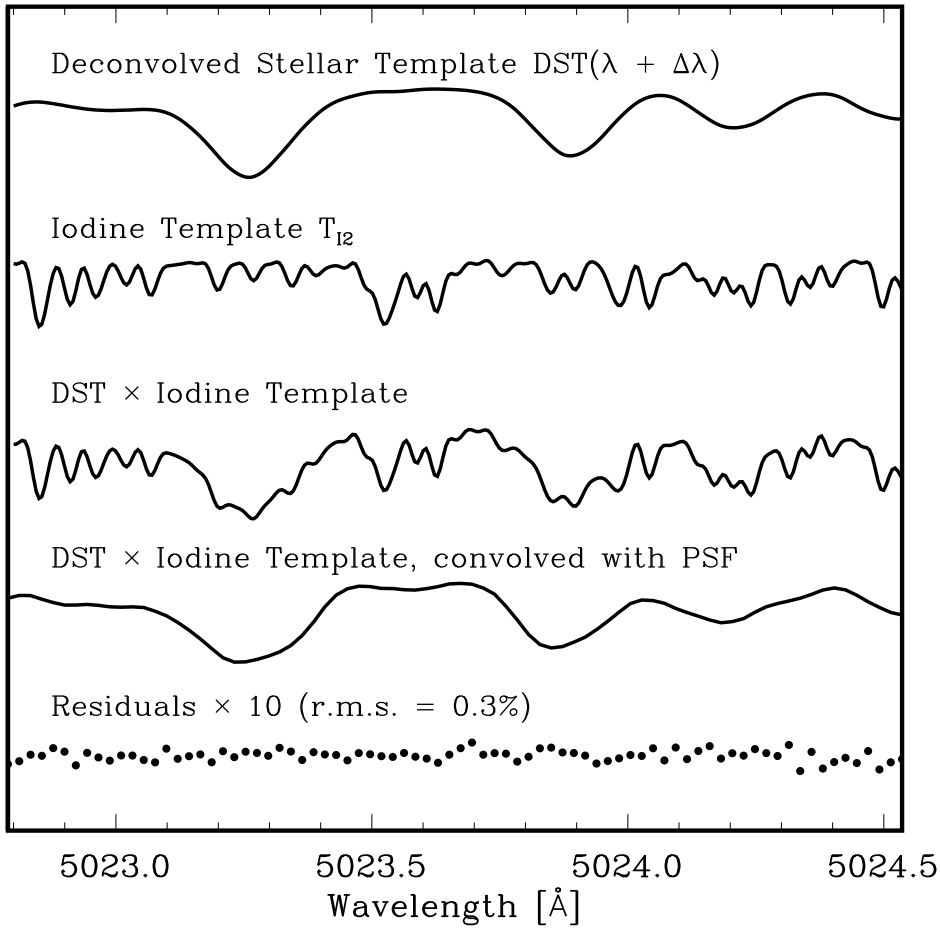}
\caption {Illustration of the iodine cell Doppler Model. From top to
  bottom, the shifted deconvolved stellar template is multiplied by
  the iodine reference spectrum. The product is then  convolved with
  the spectrometer PSF and fit to the stellar spectral observation obtained
  through the iodine cell. The residuals from the fit are shown at the
  bottom. \label{dop_example}} 
\end{figure}

Depending on the format of the echelle spectrometer, the stellar
spectra are broken into 600 to 1000 2--\AA--wide spectral chunks,
and each chunk is modeled as 

\begin{equation}
\label{iodine_model}
I_{obs}(\lambda) = k[T_{I2}(\lambda)DST(\lambda + \Delta\lambda)] * PSF,
\end{equation}

\noindent
where $k$ is a normalization factor, PSF is the model
of the spectrometer's instrumental response, and ``*'' denotes a
convolution. The fitting procedure is illustrated in
Fig.~\ref{dop_example}.
The Doppler shift, $\Delta\lambda$, is applied to the
deconvolved stellar template, DST, while the iodine reference spectrum,
$T_{I2}(\lambda)$, determines the wavelength scale of the spectral
chunk. The stellar template is multiplied by the iodine
template, and the product is convolved with the PSF and fit to the
observed spectrum. 

The fitting procedure requires two template spectra: the deconvolved
stellar template and the iodine transmission
spectrum. The iodine spectrum has been measured at high resolution 
(R $>$ 200,000) and high signal-to-noise (S/N~$>$~700) using the Fourier
Transform Spectrometer at Kitt Peak National Observatory. While the
same iodine template can be used for every RV
observation, the stellar template must be measured for each individual target
star. Stellar template observations are made without the iodine cell
in the spectrometer light path, and are obtained with higher resolution and
signal-to-noise ratios than typical iodine observations. Templates
also require bracketing iodine spectra of rapidly rotating
B stars, which are used to measure the
spectrometer PSF and wavelength scale at the approximate time of the
template observation. The PSF is then removed using a modified Jansson
deconvolution algorithm \citep{jansson}.

Our modification to the B96 reduction procedure is to derive
the DST directly from an iodine observation,
thereby circumventing the separate template observation for each
target. We use a modeling procedure similar to
Eqn.~\ref{iodine_model}, replacing the deconvolved stellar template,
DST, with a synthetic, or ``morphed,'' template spectrum. The shape
of the spectrum is varied until, when multiplied by the iodine
spectrum and convolved with the PSF, a fit is obtained to the stellar iodine
observation. This morphed deconvolved template is then substituted for
the DST in Eqn.~\ref{iodine_model} during the velocity analysis of
the target star's iodine observations.

The fitting procedure requires an accurate initial guess of the shape
of the morphed DST. 
A spectrum derived from a model stellar atmosphere would seem like an
ideal choice, since each individual spectral line could be described
analytically. But the fit of Eqn.~\ref{iodine_model} to the observed
spectrum
is highly sensitive to the exact positions of spectral lines in the
DST. Small mismatches between model and observed
spectra caused by effects such as convective blueshift can be
compensated by asymmetric adjustments to the PSF model, which can in
turn lead to spurious Doppler shifts \citep{winn05}. Modern stellar
atmosphere models cannot accurately account for the identification and
exact wavelengths of each of the  thousands of lines spanning the 1000--\AA--wide iodine region. 

\begin{figure}
\epsscale{1}
\plotone{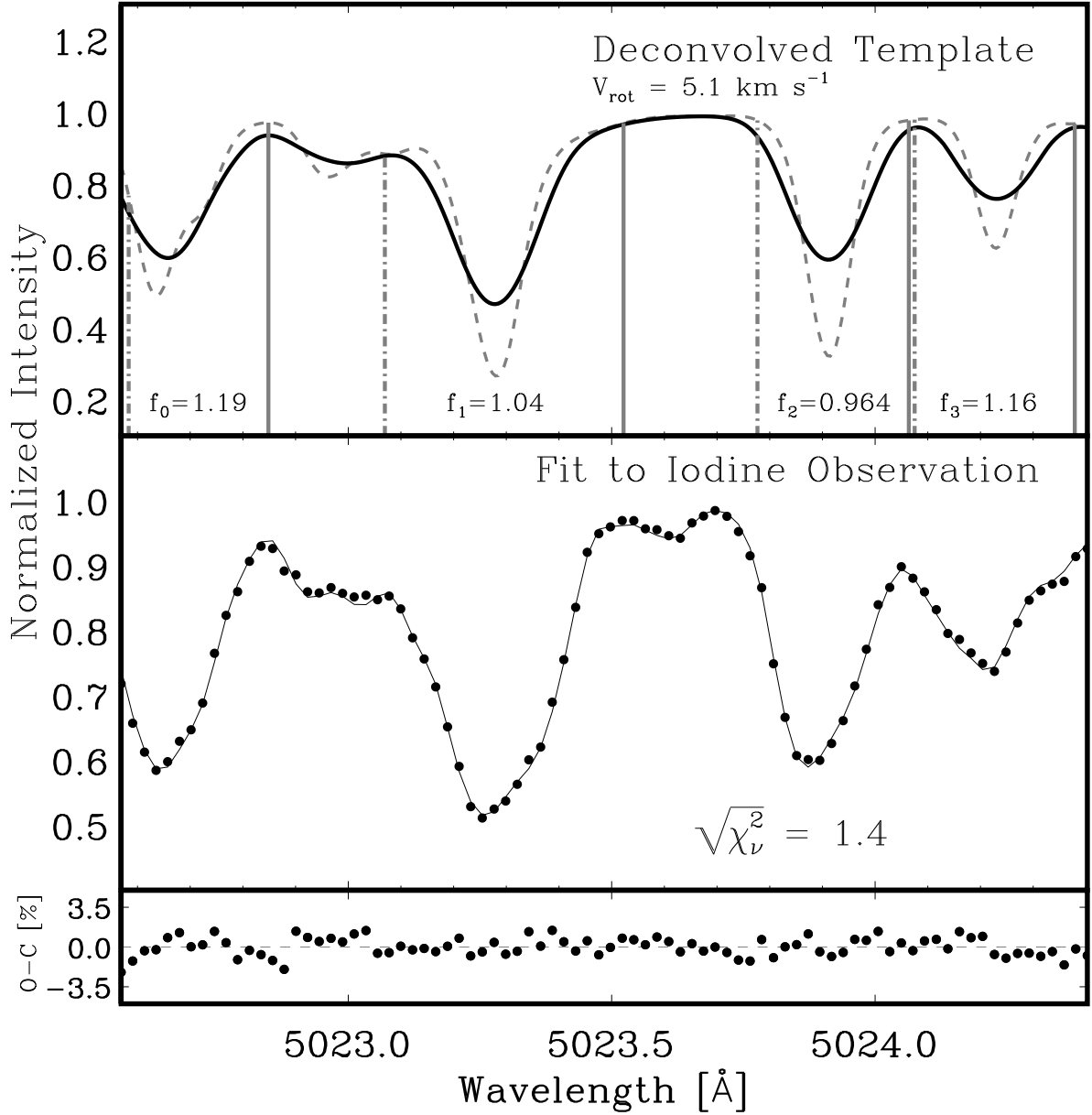}
\caption {Illustration of spectral morphing algorithm. \emph{Top}: The
  first-guess deconvolved stellar template (DST, dashed lines)
  compared to the modified synthetic DST (solid line). The synthetic
  spectrum has been rotationally broadened by 
  5.1~\ks and the four spectral segments that have been
  identified as individual lines have been
  adjusted by a pseudo-optical depth factor $f_{i}$. The borders of
  the four line segments are marked on the left (dot--dashed line) and right
  (solid). \emph{Middle}:
  The DST multiplied by the iodine reference spectrum (solid line),
  convolved by the PSF model, and shifted to fit the stellar
  observation (dots). Bottom: The fit residuals. \label{morph_example}}
\end{figure}

Instead, we find that an existing observed DST with
characteristics similar to the target star works the
best as the initial guess of the shape of morphed template.
Stars with similar surface gravity, effective temperature and
chemical composition differ only slightly in the depth and width
of individual spectral lines. We have built a library of DSTs for 42 stars
spanning the main--sequence and 
subgiant branch of the HR diagram, with Hipparcos colors $0.55 < B-V <
1.3$ \citep{hipp}, and projected
equatorial radial velocities $V_{rot}\sin{i} < 2$~\ks\
as measured by \citet{valenti05}.  The library 
templates were observed as part of the regular California \& Carnegie
Planet Search \citep[CCPS,][]{vogt00, marcy05b}, using the
Keck/HIRES echelle spectrometer with the 0.38 arcsecond wide E2 decker,
which yields a spectral resolution of $\lambda/\Delta\lambda \sim
110,000$ at 5500~\AA. Each 
library template observation consists of 3 to 5 sequential exposures, with
coadded signal--to--noise ratios ranging from 800 to 1000. The
PSF is derived from a set of bracketing B star/iodine spectra and
removed from the template spectrum using the modified Jansson
deconvolution algorithm (B96). 

The morphing procedure begins by selecting 3 to 6 library template
stars with Hipparcos colors and absolute visual magnitudes similar to
the target 
star. The non--deconvolved template spectra of the candidate library 
stars are then cross--correlated against the target star's
``pseudo--template,'' which is created by dividing the star--plus--iodine
observation by a B star/iodine spectrum from the same observing
run. The library template that yields strongest 
cross--correlation peak, averaged over all orders, is chosen as the
initial guess for the shape of the deconvolved stellar template. 

With the first--guess spectrum established, the model we use to derive
the morphed DST is

\begin{equation}
\label{morph_model}
I_{obs}(\lambda) = k[ T_{I2}(\lambda) M(DST)] * PSF
\end{equation}

\noindent
where $M(DST)$ represents a transformation of the library deconvolved
template. We do not allow the PSF to vary, but instead  
derive its shape from a B star/iodine spectrum obtained near the time
of the observation of the target star. As in Eqn.~\ref{iodine_model},
the DST is shifted by $\Delta\lambda$. But in Eqn.~\ref{morph_model}
the shape of the spectrum  
is also allowed to vary by adjusting the depths of individual lines
and globally broadening the spectrum with a
rotational broadening kernel corresponding to an equatorial rotation
velocity $V_{rot}$. 

As with the B96 method, the fit is carried out on individual
2--\AA--wide spectral chunks. However, in order to modify the line
depths of the DST, the spectral chunks are further divided into
individual line segments using an automated search algorithm. The
algorithm first locates the minima in the DST that lie further than
$4\sigma$ below the nominal continuum level, and then traces along the
line wings until either the continuum or a neighboring line segment is
encountered. Neighboring line segments corresponding to line blends
are merged and the blend is treated as 
a single line; a compromise necessary to avoid cross talk
between the two segments. 

The transformation function can be expressed as

\begin{equation}
\label{morphing}
M(DST) = R(V_{rot}) * \sum\exp\{-f_i \ln[DST_i(\lambda + \Delta\lambda)]\}
\end{equation}

\noindent
where DST$_i$ is the $i$th spectral segment of the library DST,
which is scaled by a factor $f_i$. The rotational broadening kernel,
$R(V_{rot})$, is computed with a linear limb--darkening coefficient 
appropriate for the star's $B-V$ color and the mean wavelength of the
iodine spectral region \citep{gray}. Fig.~\ref{morph_example}
shows an example of a morphed deconvolved template, together with the
resulting fit of Eqn.~\ref{morph_model} to an observed stellar iodine
observation and the fit residuals.

Since the library DST is continuum normalized, the logarithmic
term in Eqn.~\ref{morphing} can be thought of as a pseudo optical depth
under optically thin conditions. Our choice of this functional form 
is obviously not motivated by physical considerations since the
spectral lines in the template 
are not optically thin. Instead, we chose it because it has
the desirable behavior of leaving the continuum unchanged while the
depth adjustment is applied smoothly from the line wings to the
core. This has the effect of stretching the line in a somewhat elastic
fashion. To enforce continuity from one line segment to the next, each segment
has a pseudo continuum level such that portions of DST$_i$
that lie above this level are left unchanged. For a given line segment,
the pseudo continuum is the lesser of either the value at the left segment
boundary or 98\% of the stellar continuum.

In most cases, particularly with N2K targets, we derive the morphed
DST from a single iodine observation. However, since the fitting
procedure is carried  out in the rest frame of the library star, a
refined solution can be  obtained by fitting Eqn.~\ref{morph_model} to
multiple spectral observations. The composite morphed DST is formed by
replacing the line depths, $f_i$, in Eqn.~\ref{morphing} with the
weighted average,  

\begin{equation}
\overline{f_i} = \frac{\sum f_{i,j}/\sigma_j^2}{\sum1/\sigma_j^2},
\end{equation}

\noindent
where $f_{i,j}$ is the line depth adjustment for the $i$th segment of
the fit to the $j$th observation, and $\sigma_j$ is the standard deviation of
the residuals of the fit of Eqn.~\ref{morph_model}. The broadening
kernel is similarly averaged over all observations.

\section{Tests}
\label{testing}
The N2K planet search uses a quick--look observing strategy 
to efficiently search for short--period planet candidates within a 
set of $\sim2000$ target stars. Stars are typically monitored over three
consecutive nights, and those exhibiting conspicuous RV
variations---typically $>5\sigma$---are flagged for follow--up
observations. A 
Jupiter--mass planet in a short--period ($< 15$~d) orbit will
produce velocity variations of order hundreds of meters per second,
necessitating only modest velocity precision. However, the number
of extrasolar planets rises sharply with decreasing planet mass
as $dN/dM \propto M^{-1.0}$ \citep[][]{marcy05a}, which
makes it crucial that we attain the highest possible precision
in order to detect Saturn--mass planets. For example, 
a $0.3~$M$_{jup}$ planet in a 15~d orbit around a
Solar--mass star induces RV variations with an amplitude
of only 25~\ms. Reliably detecting this relatively low--amplitude signal from
three velocity measurements made during consecutive nights requires
a short--term measurement precision better than $\sim5$~\ms.

We are also interested in the long--term performance of
morphed templates. While most stars are monitored over consecutive
nights, the velocity measurements
of some stars have to be split up over several runs when telescope
time is lost to weather . Also, it would be ideal if morphed DSTs are
stable enough from one run to the next to allow for follow--up
without having to obtain a traditional template
observation. In the following subsections we compare both the short--
and long--term velocity performance of morphed DSTs to observed
templates.

\subsection{Standard Stars}

In August 2004, the Keck/HIRES spectrometer was upgraded with a new
detector. The previous 2K\ $\times$\ 2K pixel Tektronix CCD was replaced by
an array of three 4k\ $\times$\ 2k pixel MIT-LL CCDs. The new detector
produces much higher velocity precision due to its improved charge transfer
efficiency and charge diffusion characteristics; smaller resolution elements (15~$\mu m$ pixels compared to
the old 24~$\mu m$ pixels); higher quantum efficiency; increased
spectral coverage; and lower read noise. 

Since the detector upgrade, we have monitored a set of 32
chromospherically inactive stars as part of the California \& Carnegie
Planet Search in order to evaluate the RV precision of
our Doppler reduction pipeline. Each of these RV
``standard'' stars has received intensive monitoring
during at least one observing run, with multiple observations spanning
consecutive nights. Additionally, 14 of the stars have also been observed 
during at least five separate observing runs spanning more than a year
and a half. Typical signal--to--noise ratios range from 600 to more than
1000, depending on the star's apparent magnitude, with a spectral
resolution $\lambda/\Delta\lambda = 80,000$ at 5500~\AA. 

\subsection{Velocity Comparisons}
\label{vel_comparo}

\begin{figure}[t!]
\epsscale{1}
\plotone{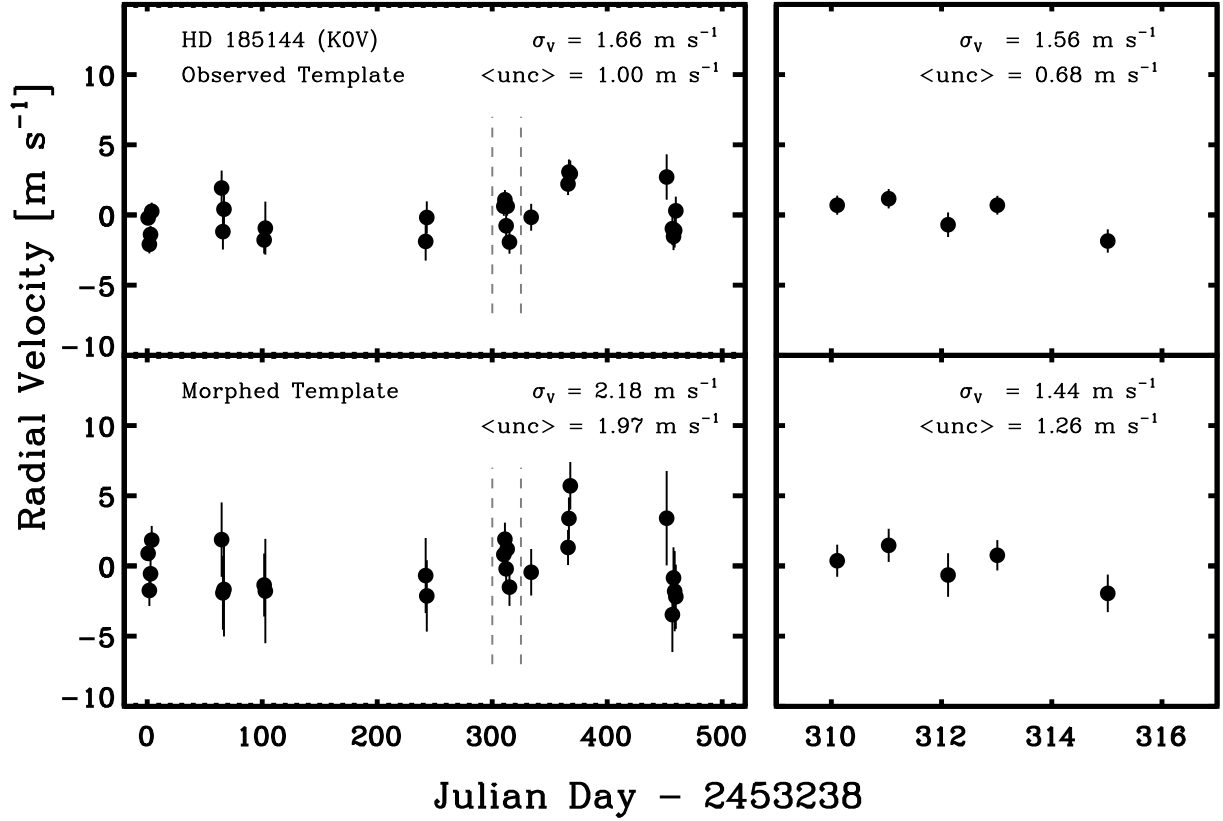}
\caption{Comparison of velocities derived from a standard observed 
  template (top) and a synthetic morphed template (bottom) for
  HD~185144 (spectral type K0V). The left panels show the long-term
64723  velocity precision for each type of template and the right panels
  show the precision during a single observing run denoted by dashed
  lines in left panels. \label{vel_comparo1}}  
\end{figure}

\begin{figure}[t!]
\epsscale{1}
\plotone{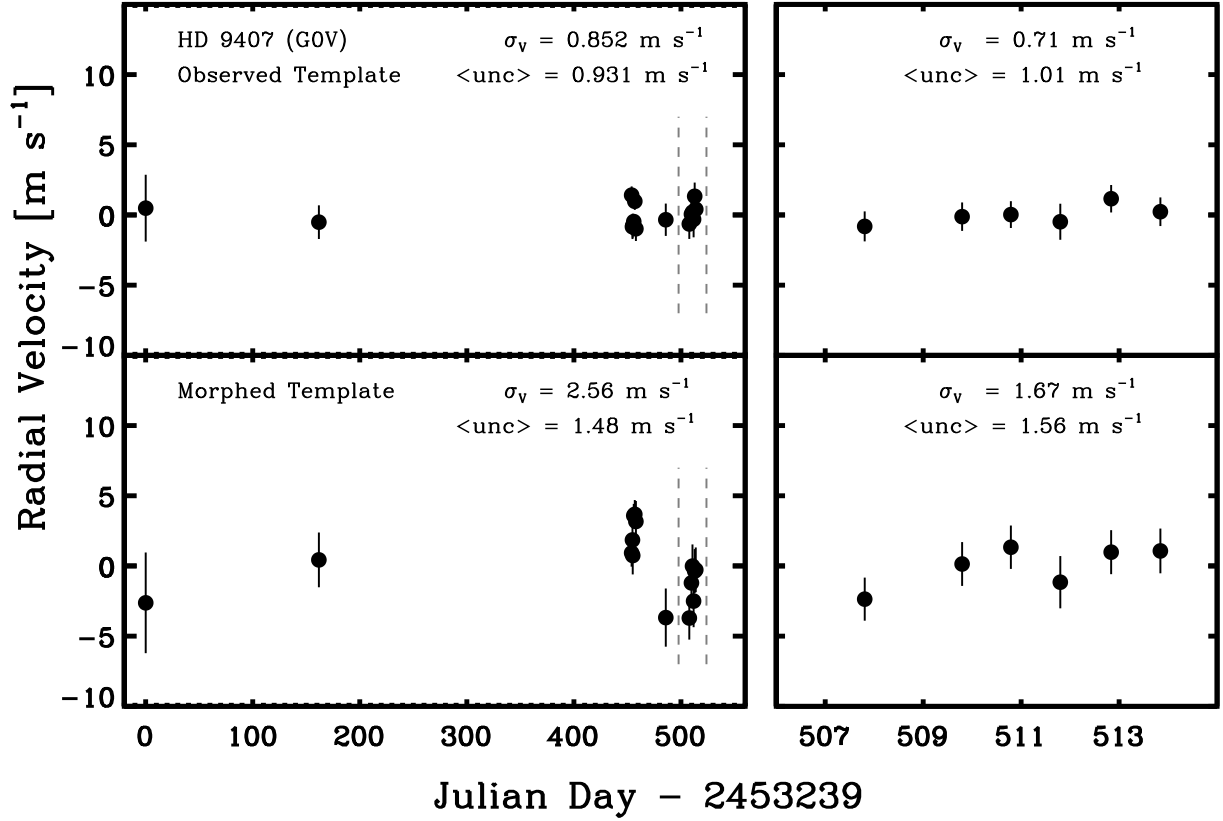}
\caption{The same as Fig.~\ref{vel_comparo1} but for HD~9407 (spectral
  type G0V).}
\label{vel_comparo2}
\end{figure}

We used these ``standard'' stars to perform a differential evaluation
of the velocity precision of morphed versus observed DSTs.
Figure \ref{vel_comparo1} shows an RV plot for one of the
more intensely observed standard stars, HD~185144 (spectral type
K0V). The top panels show the velocities measured using an observed
DST and the bottom panels show the velocities derived from a morphed
DST. The velocities measured from a morphed template exhibit a
long--term RMS scatter of 2.16~\ms, compared to 1.66~\ms~ for the
observed template. The short--term RMS for the morphed template
velocities is nearly identical to that of the observed template (1.44~\ms~
versus 1.56~\ms). Figure~\ref{vel_comparo2} shows a
similar velocity comparison for HD~9407 (G0V). In this case, the
observed--template velocities have RMS~$< 1$~\ms\ for both the
short-- and long--term test cases, while the scatters in the
morphed--template  velocities are about 2 and 3 times larger,
respectively.  

Not surprisingly, the mean internal uncertainty, $<$\emph{unc}$>$, of
the morphed velocities is larger than the internal errors produced by
the observed templates. This is likely the result of small differences 
between the morphed DST and the true stellar spectrum,
which are not adequately accounted for by Eqn.~\ref{morph_model}. 
These mismatches are also likely responsible for the systematic offsets
from run to run in the HD~9407 velocity measurements shown in
Figure~\ref{vel_comparo2} between days 450 and 550.

\begin{figure}[t!]
\epsscale{1}
\plotone{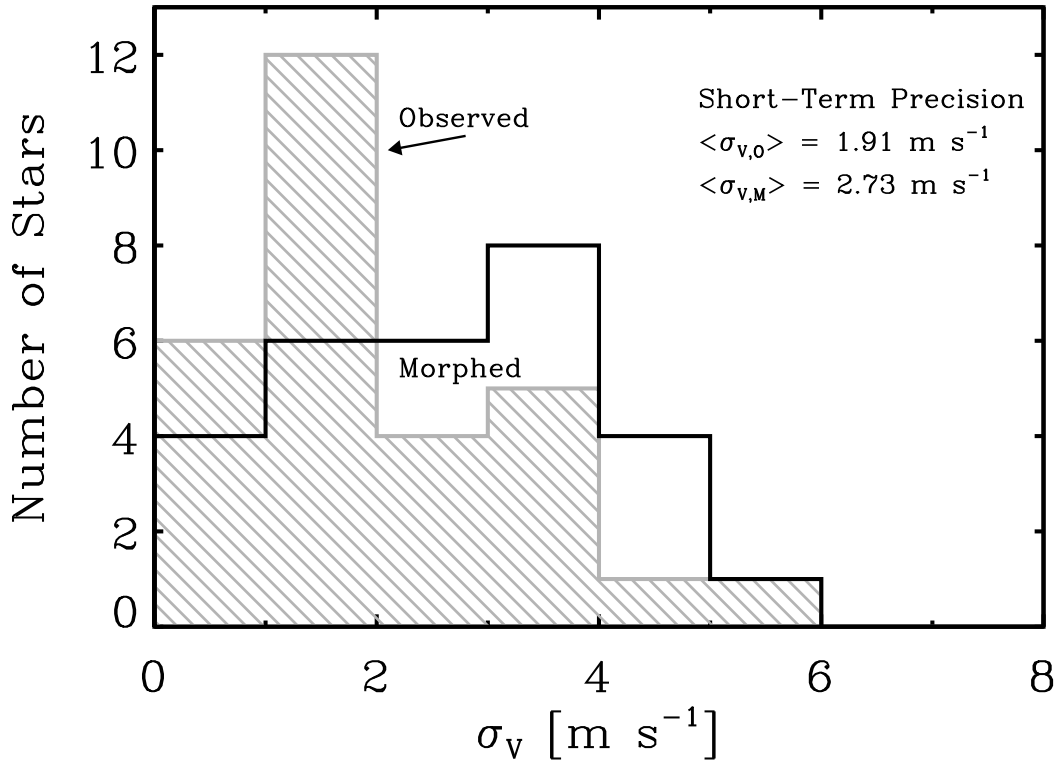}
\caption{Histograms comparing the short-term (5 night) velocity
  precision of morphed  versus observed (shaded)
  templates for a set of 32 chromospherically quiet radial velocity
  standards. The mean RMS scatter of the velocities derived
  from morphed and observed templates is given by  $\langle\sigma_{V,M}\rangle$ and $\langle\sigma_{V,M}\rangle$, respectively.}
\label{shortterm}
\end{figure}

\begin{figure}[t!]
\epsscale{1}
\plotone{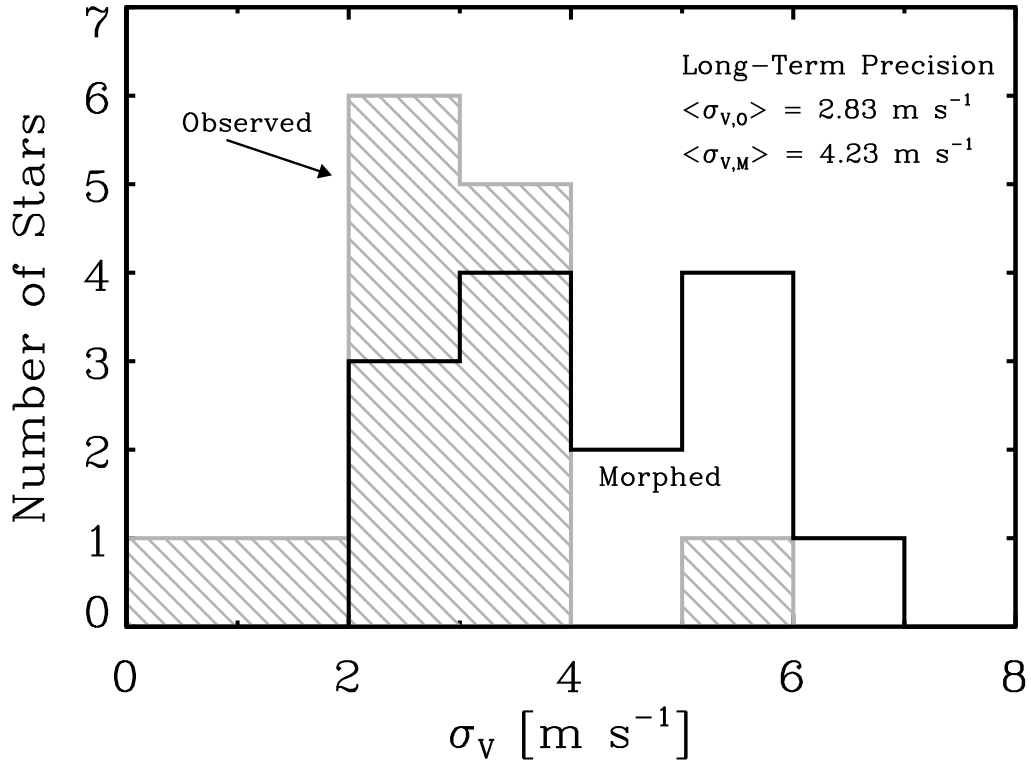}
\caption{The same as Fig.~\ref{shortterm} but for a subset of 14
  standard stars observed during more than five observing runs 
  spanning $\sim1.5$ years.}
\label{longterm}
\end{figure}

Figure~\ref{shortterm} shows histograms of the short--term (single
run) velocity scatter, $\sigma_{V}$, measured for the ensemble 
set of 32 standard stars using morphed and
observed templates. Figure~\ref{longterm} shows a similar
comparison for the 14 stars with $\sim1.5$ years of observations
spanning more than five separate observing runs. A comparison of the
mean velocity scatter shows that morphed templates perform within a
factor of 2 of observed templates, with a typical precision of
$\sim3$~\ms~within a single observing run and a long--term precision
of $\sim5$~\ms.

It should be cautioned that Figures~\ref{shortterm} and~\ref{longterm}
do not represent unbiased tests of the absolute precision attainable
with the B96 technique. Such a test is complicated by differences in
the quality of observations from one run to the next, the time
coverage and frequency of the observations, and the possible presence of
undetected low--mass planets. Instead, our tests 
represent a differential comparison of the precision produced by
morphed and observed templates. From this test we can conclude that
morphed templates are well--suited to the needs of the N2K
observing program, providing the necessary short--term precision 
for the initial detection Saturn--mass planets, and long--term precision
adequate for monitoring planet candidates from one run to the next.

\section{Application to N2K}
\label{application}

During the 2005B Keck observing semester, we observed 63 metal--rich
stars chosen based on the criteria described in Paper I 
\citep[see also][]{robinson06, ammons06}, using the upgraded HIRES
spectrometer. We measured velocities from the spectral observations
using the B96 Doppler technique, with morphed DSTs created
with the first iodine observation of each star.

\begin{figure}[t!]
\epsscale{0.9}
\plotone{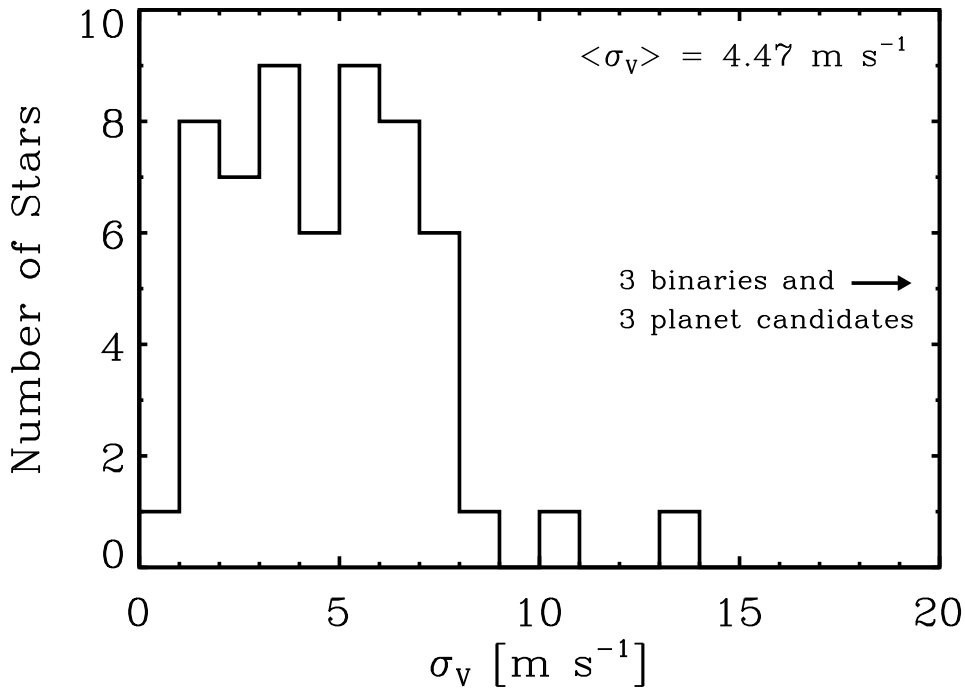}
\caption{Distribution of RMS velocities measured for 64 N2K
  target stars. Three binaries (SB1) and three
  planet candidates with $20 < \sigma_V < 250$~\ms~lie outside of
  the plot range. The three stars with $\sigma_V > 9$~\ms~ have jitter
  estimates in excess of 10~\ms. With these 8 outliers removed, the mean
  is $\langle\sigma_V\rangle = $~\nksig~\ms, with 87~\% of the 
  stars having $\sigma_V < 10$~\ms.  \label{j13_hist}}
\end{figure}

Figure \ref{j13_hist} shows a histogram of the velocity RMS
for our sample. For stars with linear trends and slopes
greater than 10~\ms~d$^{-1}$, the scatter is measured with respect
to a linear fit to the three data points. The majority of the stars
(87\%) had RMS~$< 10$~\ms. The two stars with
velocity scatter between 10 and 20~\ms~ had predicted velocity
``jitter'' values in excess of 10~\ms~ due to large chromospheric
activity indices \citep{wright05}. Three stars had
$\sigma_V > 250$~\ms, 
with linear trends indicative of a stellar--mass orbital companion,
and three stars showed evidence of a short--period planet, with $50 <
\sigma_V <  250$.  We obtained additional observations of the three planet 
candidates at Keck Observatory, and present here the detection of a
1.5~\mjup~ planet in a 2.1~d orbit around HD~86081. The remaining
candidates are still being monitored at Keck. 

Excluding the 8 stars showing excess
variability, the mean velocity RMS of our sample is
\nksig~\ms. The mean RMS
for our N2K targets is larger than the short--term test case shown in
Fig. \ref{shortterm} due to the lower signal-to-noise of the
observations (S/N~$\lesssim 200$). 

However, this performance represents a marked improvement 
over the previous synthetic template technique used in Paper
I, which used the NSO solar atlas in lieu of our current library
of DSTs. Our old method produced a mean velocity scatter of $\sim15$~\ms~
and a long tail in the distribution extending to 60~\ms. 
Of the original sample of 211 N2K targets monitored at Keck beginning in
2004 and analyzed using our older method, 35 were flagged as planet
candidates based on having RMS scatter $> 3\sigma$ (Paper I). These
stars received additional observations, and we performed the Doppler analysis
using observed templates to confirm their
variability. Three planets from this original sample have been
reported previously \citep[Paper I,][]{fischer06}. We report here
the detection of two new planets from our original sample of Keck
targets: a $0.72$~\mjup~ planet orbiting HD~224693 with circular
26.7~d period and a Saturn--mass planet in an eccentric 18.2~d orbit
around HD~33283.   

In the following sections we present the stellar parameters of the
host stars, orbital parameters of the planets, and the result of our
search for transit events. 

\subsection{HD~86081}
\label{86081}
HD 86081 is an \spa\ star with $V = \vmaga$, $B - V = \bva$, a Hipparcos 
parallax--based distance of \da\ parsecs \citep{hipp}, and an absolute 
visual magnitude, $M_V = \mva$.  The star is chromospherically 
inactive, with no emission seen in the core of the Ca~II~H~\&~K lines. 
We measure $S_{HK} = \shka$ and $\log R'_{HK} = \rhka$ 
and derive a rotational period of \pra~d following the calibration 
of \citet{noyes84}.  Our high resolution spectroscopic analysis, 
described in \citet{valenti05}, yields $T_{eff} = \teffa$~K, 
\logg~$=\ga$, \fe~$=+\fea$ and \vsini~$ = \va$~\ks. Interpolation
using the \citet[][, hereafter Y$^2$]{yi02} isochrones 
provided a stellar mass estimate of $\msa \pm 0.05$~\msun\ and a stellar
radius of $\rsa \pm 0.1$~\rsun. The stellar parameters are summarized
in Table~\ref{stars}. 

\begin{figure}[t!]
\epsscale{0.9}
\plotone{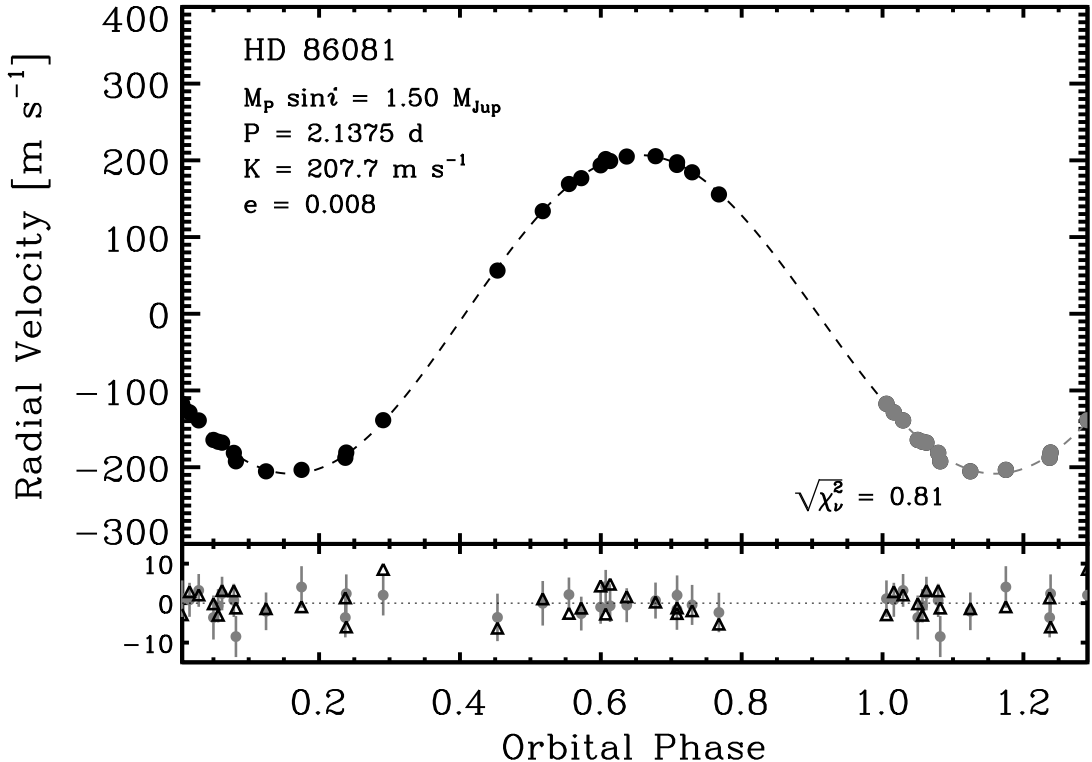}
\caption{Top: Phased radial velocity measurements of HD~86081
  (observed template). The dashed line shows the best--fit orbital
  solution. Bottom: The residuals of the orbital fit (filled circles)
  and the difference between the velocities measured with observed and
  morphed templates (open triangles). The RMS of the fit residuals is
  \rmsa~\ms~ and the RMS difference between the two sets of RV
  measurements is 2.8~\ms. \label{orbit86081}}
\end{figure}

We started monitoring HD~86081 in 2005 November, and our Doppler
analysis using a morphed template yielded a non--linear velocity
variations with RMS~$ = 154$~\ms\ from the first three
measurements. Additional monitoring revealed a 
clear, \pa~d periodicity. We list our velocity measurements in 
Table~\ref{vel86081} along with the Julian Dates and internal
measurement uncertainties. We report here only the velocities measured
using a high--quality, observed template, due to the higher velocity
precision.

In order to search for a the best--fit orbital solution, we augmented
the internal measurements errors by adding in quadrature a photometric 
jitter estimate of 3.7~\ms\ based on the  chromospheric activity index
of the star \citep{wright05}. Figure~\ref{orbit86081}
shows the phased RV observations of HD~86081 together with the
best--fit orbit solution: $P = $~\pa~d, $K = $~\ka~\ks, and 
$e = $~\ea\plmn\eea. From the
orbital parameters and adopted stellar mass of \msa~\msun, we
derive a minimum planet mass \msini~$ = \ma$~\mjup\ and semimajor axis
$a = \aa$~AU. The lower panel of
Fig.~\ref{orbit86081} shows the fit residuals (filled circles), which have an
RMS scatter of \rmsa~\ms\ and \chisq~$ = \chia$. Also shown in the
lower panel are the differences between the measurements made with morphed 
and observed templates (open triangles), which show excellent
agreement (RMS~$=2.8$~\ms). 

We list the best--fit orbital parameters in Table~\ref{param86081}
along with the parameters determined from the morphed--template
velocities. The two sets of parameters are identical within errors.
The parameter uncertainties were estimated using a Monte
Carlo method. For each of 1000 trials, the fit
was subtracted from the measurements, the scrambled residuals were
added back to the measurements, and a new solution was
determined. The standard deviations of the parameters derived from all
trials were adopted as the 1$\sigma$ uncertainties.

\subsection{HD~224693}

\begin{figure}[t!]
\epsscale{0.9}
\plotone{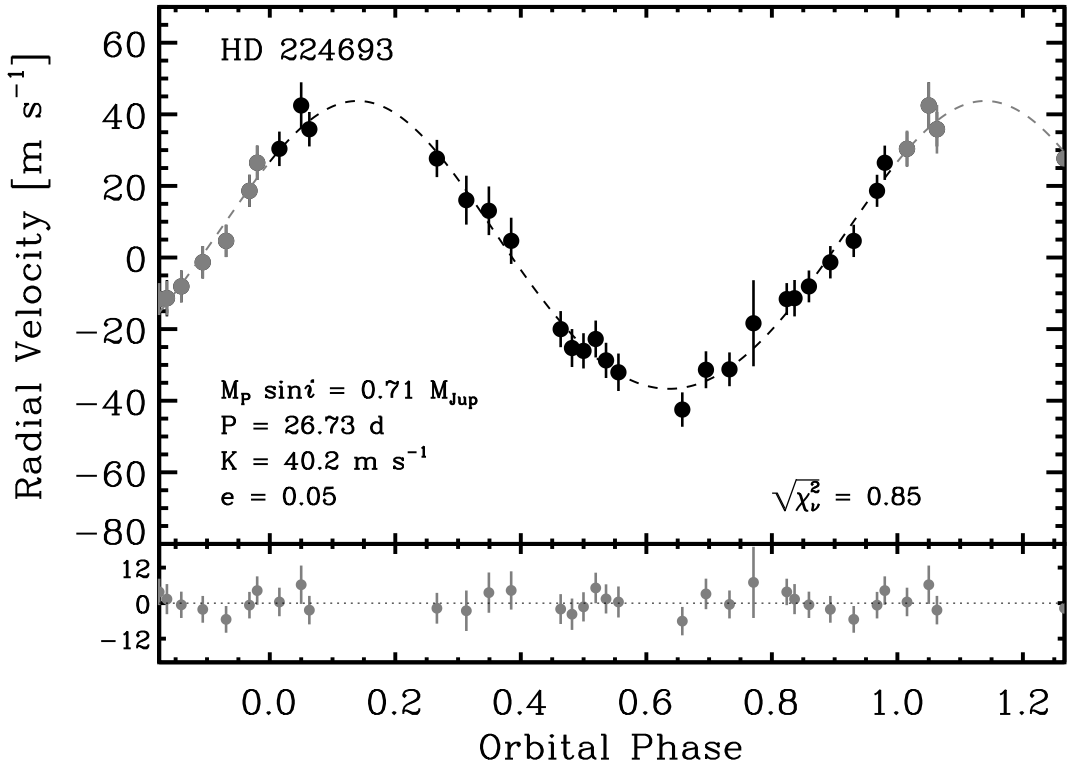}
\caption{Phased radial velocity mesurements of HD~224693 and best--fit
  orbital solution. The fit residuals have an RMS of \rmsc~\ms.
 \label{orbit224693}} 
\end{figure}

HD 224693 is a $V = \vmagc$, \spc\ star with $B-V = \bvc$, a
parallax--based distance of 94~pc and $M_V = \mvc$ \citep{hipp}. The
star is chromospherically quiet, with $\log R'_{HK} = \rhkc$ and a
rotational period of \prc~d, based on  the calibration by
\citet{noyes84}. Our spectral synthesis modeling determines 
$T_{eff} = \teffc$~K, \logg~$ = \gc$, \vsini~$ = \vc$~\ks and \fe$ =
+\fec$.  The mass of the star  
derived from Y$^2$ isochrones \citet{yi02} is estimated to be
\msc\plmn0.1~\msun\ with a radius  of \rsc\plmn0.3~\rsun. 

HD~224693 was observed at Keck as part of the N2K program beginning in
2004 July. The first four Doppler observations showed a modest
velocity RMS of 16~\ms. 
At that time, the synthetic template algorithm was still being tested,
so a standard template was made to confirm the low amplitude variation. 
We obtained a total of \nobsc\ radial velocities, listed in
Table~\ref{vel224693}. We augmented the internal uncertainties with a
jitter estimate of 4.0~\ms~ based on its chromospheric activity index
\citep{wright05}. 

\begin{figure}[t!]
\epsscale{0.9}
\plotone{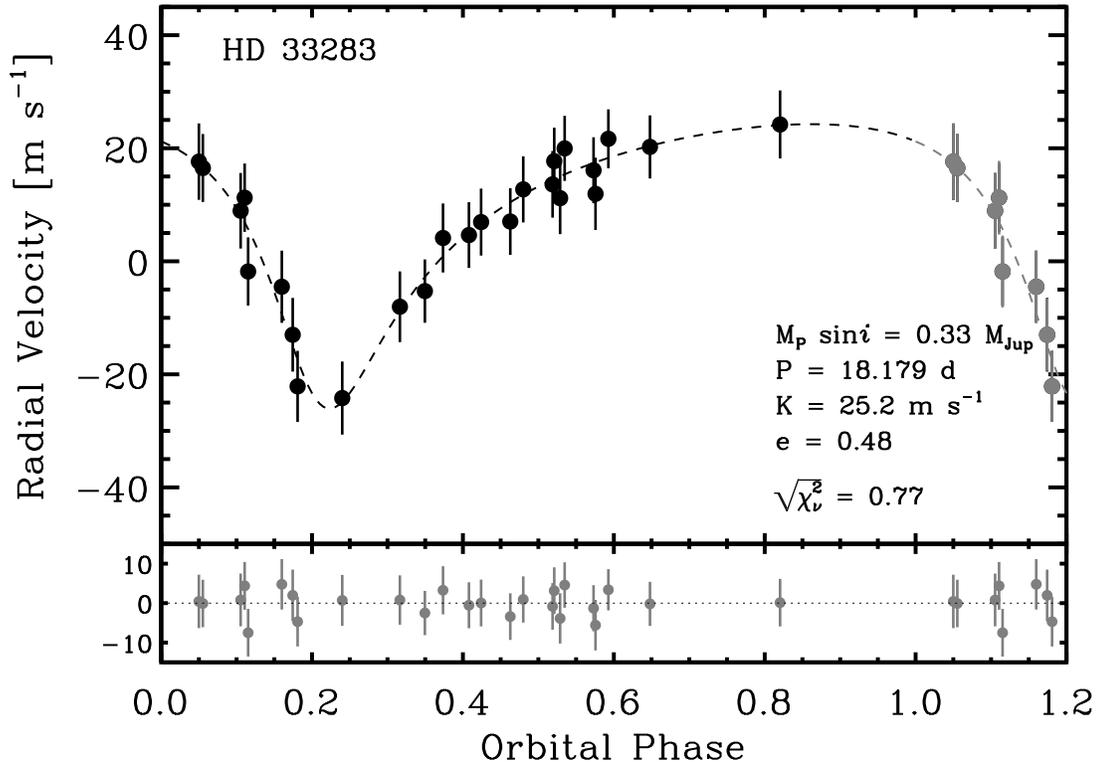}
\caption{Phased radial velocity mesurements of HD~33283 and best--fit
 orbital solution. The fit residuals have an RMS of \rmsd~\ms.
 \label{orbit33283}} 
\end{figure}

The velocities are well fit by a Keplerian with $P =
\pc$~d, $K = \kc$~\ms,  and eccentricity,
$e = \ec$\plmn\eec.   Figure~\ref{orbit224693} shows the phased radial velocity
measurements along with the best--fit Keplerian model. The RMS fit to
the model is \rmsc~\ms\  with \chisq~$ = $~\chic, consistent with the
measurement errors. From our
orbital parameters and the adopted  
stellar  mass of \msc~\msun\ we derive a minimum planet mass \msini~$ =
$~\mc~\mjup\ and semimajor axis $a = \ac$~AU. The full
list of derived orbital parameters is given in Table~\ref{paramothers}.

\subsection{HD~33283}

HD~33283 is a $V = \vmagd$, \spd\ star with $B-V = \bvd$.  Based on the
Hipparcos  
parallax, it has a distance of \dd~pc and absolute visual magnitude
$M_V = \mvd$ \citep{hipp}. The star is chromospherically inactive with
$S = \shkd$, $\log R'_{HK} = \rhkd$ and a predicted rotational period of
\prd~d. Spectral synthesis modeling yields $T_{eff} = \teffd$~K,
\logg~$ = \gd$, \vsini~$ = \vd$~\ks, \fe~$=+\fed$. Interpolation of
the Y$^2$ \citep{yi02} isochrones 
suggests a mass of \msd\plmn0.1~\msun\ and radius of
\rsd\plmn0.1~\rsun. We summarize the stellar characteristics in
Table~\ref{stars}.  

We began monitoring HD~33283 at Keck in 2004 July. The first five
radial velocities showed a modest RMS of 
21~\ms, with a 17~\ms\ change between the first and second observation.
At that time, our synthetic template algorithm was still
being tested, so a standard observed template was obtained to confirm
the variations. The \nobsd\ velocities measured with the observed
template are listed in Table~\ref{vel33283} along with the internal
measurement uncertainties. Based on the chromospheric emission index,
we expect 4.0~\ms\ of stellar jitter, which we added in quadrature to the
internal uncertainties. 

The best--fit Keplerian orbit solution yields $K = \kd$~\ms, $P =
\pd$~d, and $e = \ed$. The phased radial velocities and orbit solution
are show in Figure~\ref{orbit33283}. The fit has RMS~$ = $~\rmsd~\ms\ and
\chisq~$ = $~\chid, consistent with the measurement errors. From our
adopted stellar mass and best--fit orbit parameters, we estimate a
minimum planet mass of \md~\mjup\ and $a = \ad$~AU. The full orbit
solution is summarized in Table~\ref{paramothers}.

\subsection{Photometry of the Host Stars}
\label{photometry}
Between 2005 November and 2006 February, we obtained high--precision 
photometry of two of the three planetary host stars in this paper with the 
T8 and T10 0.8~m automatic photometric telescopes (APTs) at Fairborn 
Observatory.  The APTs can detect short-term, low-amplitude brightness 
variability in the stars due to rotational modulation in the visibility of 
magnetic surface features such as spots and plages \citep[e.g.,][]{henry95} 
as well as longer-term variations associated with stellar magnetic cycles 
\citep{henry99}.  The photometric observations can help to establish whether 
observed radial velocity variations are caused by stellar activity or 
planetary-reflex motion \citep[e.g.,][]{henry00a}.  \citet{queloz01} and 
\citet{paulson04} have found several examples of periodic radial velocity 
variations in solar-type stars caused by photospheric spots and plages.  
The APT observations are also useful to search for possible transits of the 
planetary companions \citep[e.g.,][]{henry00b,sato05}.

The T8 and T10 APTs are both equipped with two-channel precision 
photometers each employing two EMI 9124QB bi-alkali photomultiplier tubes to 
make simultaneous measurements in the Str\"omgren $b$ and $y$ passbands.  
The APTs measure the difference in brightness between a program star and a 
nearby comparison star.  The local comparison stars used for the program 
stars were HD~84664 and HD~203 for HD~86081 and HD~224693, respectively.
Str\"omgren $b$ and $y$ differential magnitudes were  
computed and corrected for differential extinction with nightly extinction 
coefficients and transformed to the Str\"omgren system with yearly mean 
transformation coefficients.  The external precision of the differential 
magnitudes is typically between 0.0012 and 0.0017 mag for these telescopes, 
as determined from observations of pairs of constant stars.  Finally, we 
combined the Str\"omgren $b$ and $y$ differential magnitudes into a single 
$(b+y)/2$ passband to maximize the precision of the photometric measurements.  
The individual photometric observations of HD~86081 and HD 224693 are
given in Tables~\ref{photmeas}A and \ref{photmeas}B, respectively.
Further information on the automatic telescopes, photometers,
observing procedures, and data reduction techniques can be found in
\citet{henry99} and \citet{eaton03}.    

\begin{figure}
\epsscale{0.85}
\plotone{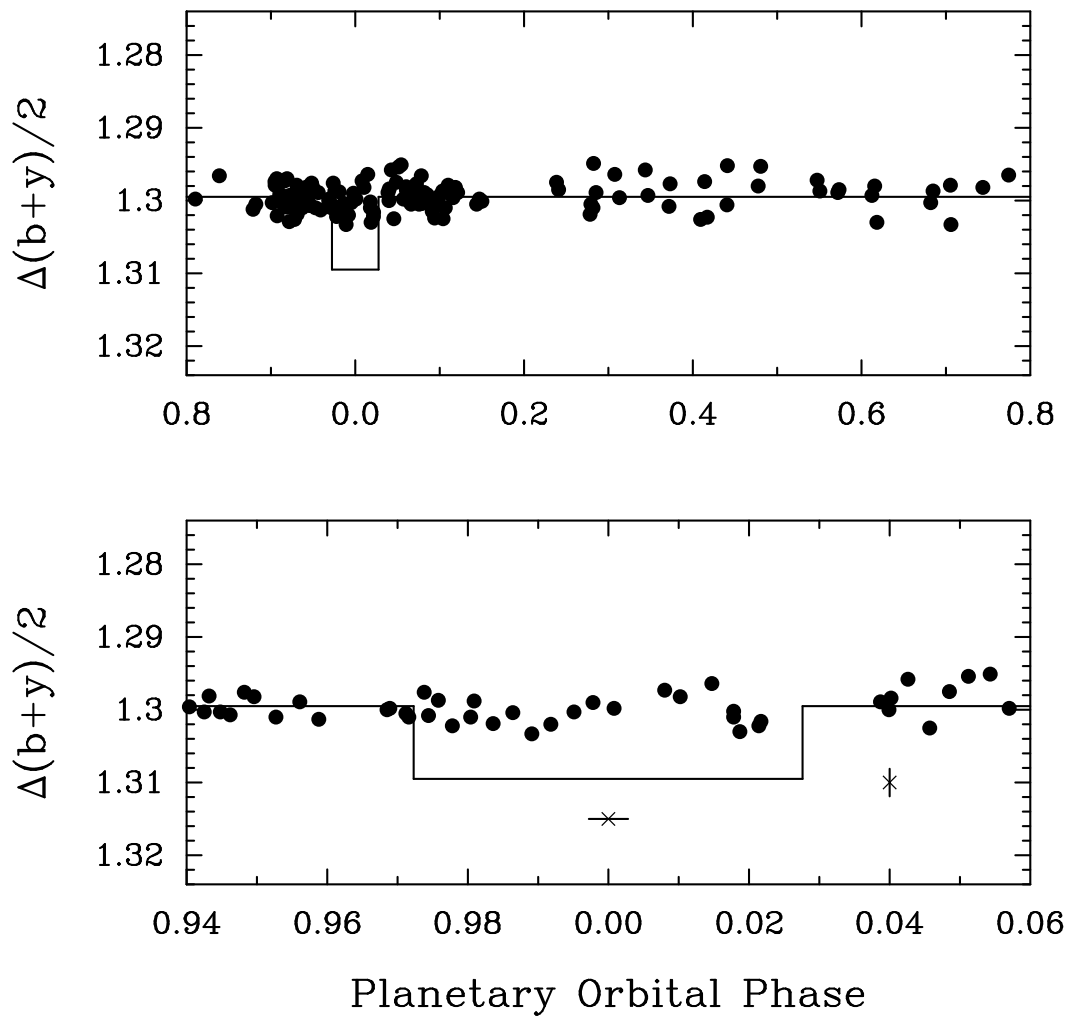}
\figcaption{{\it Top}:  The complete set of Str\"omgren $(b+y)/2$ photometric 
observations of HD~86081 obtained with the T8 0.8~m APT at Fairborn 
Observatory and plotted against orbital phase of the planetary companion.
The predicted time, depth, and duration of possible transits are shown
schematically.  The star exhibits no optical variability on the radial
velocity period larger than 0.0004 mag or so.  {\it Bottom}:  The
observations around the predicted time of transit are replotted with
an expanded scale on the abscissa.  The error bars are described in the
text.  Transits deeper than 0.001 mag or so are ruled out by these
observations.\label{lc1}}
\end{figure}

\begin{figure}
\epsscale{0.85}
\plotone{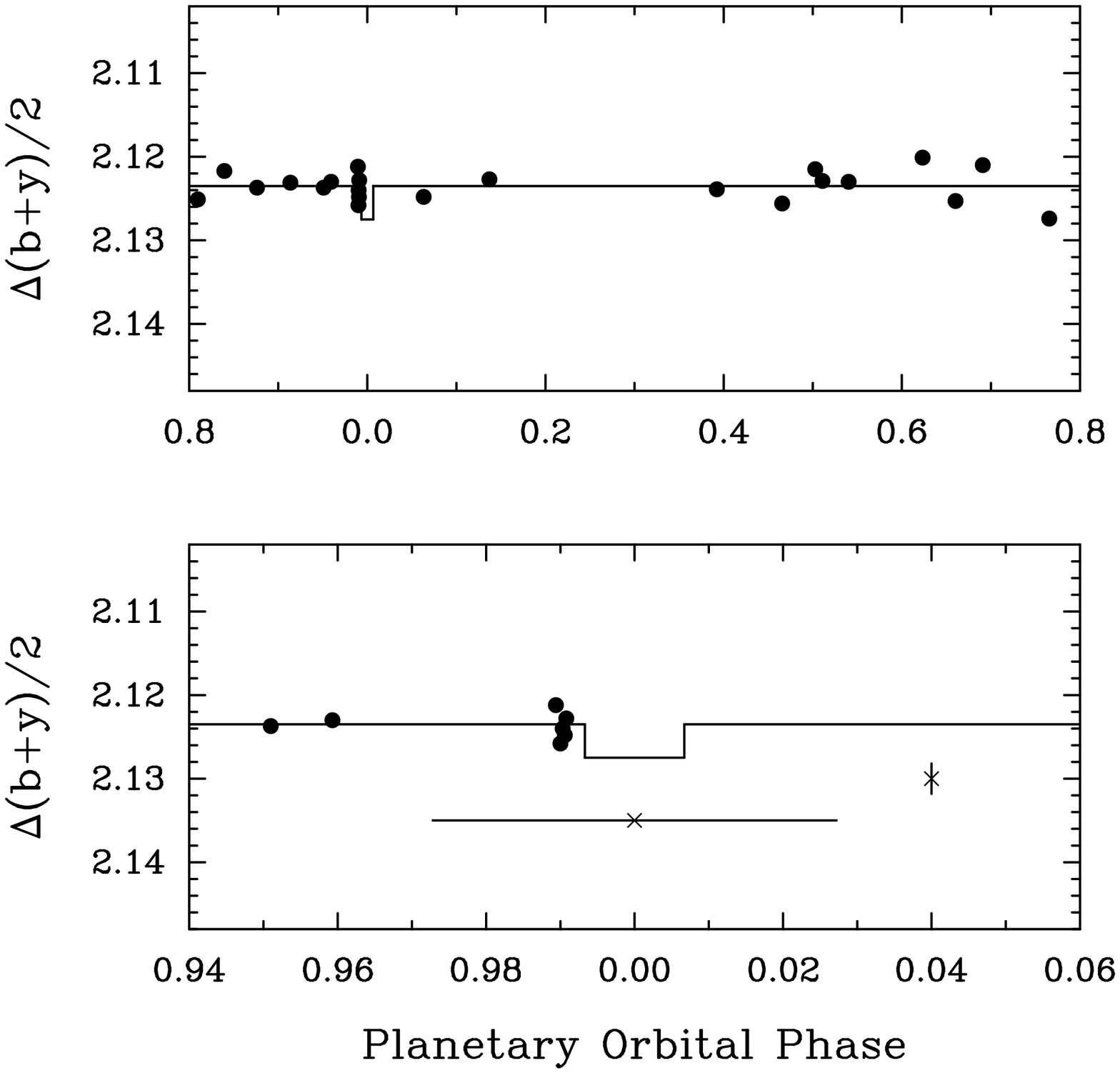}
\figcaption{{\it Top}:  The complete set of Stro\"mgren $(b+y)/2$ photometric
observations of HD~224693 obtained with the T10 0.8~m APT at Fairborn
Observatory and plotted against orbital phase of the planetary companion.
The predicted time, depth, and duration of possible transits are shown
schematically.  The star exhibits no optical variability on the radial
velocity period larger than 0.0003 mag or so.  {\it Bottom}:  The
observations around the predicted time of transit are replotted with
an expanded scale on the abscissa.  The error bars are described in the
text.  Our observations are too few to rule out possible transits.\label{lc2}}
\end{figure}

The photometric results for the two stars we measured are summarized in 
Table~\ref{photres}.  We have no photometry as yet on HD~33283.  Columns
2 and 3 give  
the date range of the photometric observations and the total number of 
individual differential magnitudes, respectively.  The standard deviations 
in column 4 refer to the spread of the $(b+y)/2$ measurements around the mean 
values of the data sets.  All standard deviations are consistent with the 
constancy of the host stars and their comparison stars.  
Periodogram analyses of the two data sets do not reveal any significant 
periodicities.  We computed the semi-amplitudes of the light curves (column~5) 
with least-squares sine fits of the observations phased to the radial 
velocity periods.  The resulting amplitudes are both 0.0004 mag or below;  
this absence of light variability on the radial velocity periods supports 
planetary-reflex motion as the cause of the radial velocity variations in 
both stars.  Although we have no photometry of HD~33283, it is the 
most inactive of the three stars (Table~\ref{stars}) and so should show
a similar lack of photometric variability.

The sixth column of Table~\ref{photres} gives the geometric probability
of transits for  
the three systems we observed photometrically, computed from equation 1 of 
\citet{seagroves03} and assuming random orbital inclinations.  The final column
gives the results of our search for planetary transits of the two
companions as shown in Figures \ref{lc1} and \ref{lc2} and discussed below.

The top panel in each of Figures~\ref{lc1} and \ref{lc2} plots the complete
photometric data set for each star against the planetary orbital
phase, computed from the  
orbital elements in Tables~\ref{param86081} and \ref{paramothers}; zero
phase in each case refers to a time  of inferior conjunction (mid--transit).  The solid curve in each panel  
approximates the predicted transit light curve, assuming a planetary orbital 
inclination of 90\arcdeg (central transits).  The out--of--transit light level 
corresponds to the mean brightness of the observations.  The transit 
durations are calculated from the orbital elements, while the transit depths 
are derived from the estimated stellar radii (Table~1) and planetary radii 
computed with the models of \citet{bodenheimer03}.  The bottom panel
of each figure  
shows the observations around the predicted time of mid transit replotted 
with an expanded scale on the abscissa.  The horizontal error bar below 
each predicted transit curve represents the approximate $\pm 1 \sigma$ 
uncertainty in the time of mid transit, based on Monte Carlo simulations 
and the uncertainties in the orbital elements.  The vertical error bar 
represents the precision of the measurements in each case.

Our photometric observations of HD~86081 cover the predicted transit window 
quite well.  The 21 observations within the transit window in the bottom panel
of Figure~\ref{lc1} have a mean brightness level of 1.3003 $\pm$ 0.0004 mag.  
The 114 observations outside the transit window have a mean of 1.2993 
$\pm$ 0.0002 mag.  Thus, the two values agree to within 0.001 mag, or
$\pm 2 \sigma$, placing very shallow limits on possible transits.  In the
case of HD~224693, our data around the predicted time
of transit are far to sparse to rule out possible transits at this time
(Figure~\ref{lc2}) and we are continuing to monitor this star.

\section{Summary and Discussion}
\label{discussion}

We have presented a modification to the B96 precision radial
velocity technique that eliminates the need to obtain template
observations for each target star. We replace observed templates with
synthetic versions derived by modifying the 
deconvolved stellar template (DST) of a star with similar
characteristics as 
the target star. The depths and widths of the spectral lines in the
surrogate DST are modified until, when multiplied by an iodine
template and convolved with a model of the spectrometer PSF, a fit
to the target star's observed star--plus--iodine spectrum is obtained.
Based on our tests of Keck/HIRES spectra of a set of
chromospherically inactive stars, we have
determined that morphed templates yield velocity measurements with
precision within a factor of two of those produced by observed
templates; with a short--term precision of $\sim3$~\ms\
and long--term stability within 5~\ms.

While observed templates yield the highest velocity precision, the
observational time savings afforded by morphed
DSTs make them favorable alternatives for many types of planet search
programs. Examples include surveys containing evolved stars such as
K giants, where stellar jitter is the dominant noise source
\citep{frink02, sato05b};
surveys conducted on small telescopes where it is not possible to
obtain high enough signal--to--noise for template observations;
testing for binarity among SIM grid star candidates
\citep{frink01,gould01}; and surveys containing faint stars such
as M dwarfs.

Morphed templates are also ideal for the ``Next
2000 Stars'' (N2K) planet search, which uses a quick--look
observing strategy to efficiently search for short--period planets within
a large sample of stars (Paper I). Synthetic templates are an integral
part of the N2K observing strategy and have already yielded important 
results.  We report here the discovery of three new
planets from the Keck N2K program: a \ma~\mjup\ planet in a \pa~d
orbit around HD~86081; a Saturn--mass planet in an
eccentric, \pd~d orbit around HD~33283; and a Jovian--mass planet with a
circular, \pc~d orbit around HD~224693.

The quick--look strategy used by the N2K consortium involves observing
stars during three consecutive nights, and Monte Carlo simulations
have shown that the strategy is most sensitive to planets with \msini~$ >
0.5$~\mjup\ and periods between 1.2 and 14 days (Paper I). However, the
planets orbiting HD~33283 and HD~224693 have orbital periods
significantly longer than 14 days. These detections were aided by the
fact that intensive follow--up observations were not initiated until
after the fourth or fifth observations were obtained. These 
two cases serve as important illustrations of how the
sensitivity of the N2K quick--look observing strategy can be greatly
enhanced with additional observations and an increased time baseline.

The \pa~d orbital period of HD~86081b bridges the gap between the
 3 day pile--up seen in the 
period distribution of hot jupiters \citep[HJ, $3 \lesssim P <
  15$~d;][]{butler06, udry03} discovered by RV surveys,
and the extremely short period ($P\lesssim2$~d) planets found by the Optical
Gravitational Lensing Experiment planet transit search
\citep[OGLE;][]{konacki03,bouchy04}. While these ``very hot jupiters''
(VHJ) are common amongst the OGLE detections, comprising 3 out of 5
transit detections, Doppler surveys have previously found only two other
Jovian--mass planets with a period significantly less than 3 days
\citep[HD~73256, $P=2.55$~d and HD~189733,
  $P=2.22$~d][]{udry03,bouchy05}. \citet[][hereafter G05]{gaudi05} showed that this apparent 
inconsistency between RV and transit searches can be reconciled by 
considering the different selection biases for the two search methods
and the $P^{-5/2}$ sensitivity scaling for ground--based transit
searches. By accounting for these effects, G05
concluded that VHJs represent a rare class of exoplanet, with an occurrence
rate only 10\% that of HJs, based on the aggregate sample from RV and
transit searches.

Since RV surveys such as the California \& Carnegie Planet Search (CCPS)
and N2K are complete down to Saturn--mass planets with orbital periods
less than 15 days ($K = 25$~\ms), we can use the combined sample of
planets from these two surveys to estimate the relative
number of VHJs and HJs. Out of $\sim1800$ total stars, these two
RV surveys have yielded 16 HJs and two VHJs (using a somewhat arbitrary
lower cutoff at $P=3$~d and therefore including HD~83443 as done
by G05). Thus, we estimate the ratio of HJs to VHJs
is $\sim8$, which is in quantitative agreement with the
estimates of G05. A much more detailed analysis will be provided in a
future publication in the N2K Consortium series \citep{fischer06b}.

The relative paucity of VHJs provides an interesting puzzle, and the
answer is likely related to the way in which close--in giant planets
are formed and how they interacted with their early circumstellar
environments. One of the 
leading theories suggests that giant planets migrate inward along
nearly circular orbits through interactions with their
circumstellar disks. These interactions can take the form of
gravitational scattering and physical collisions with planetesimals in
the remnant disk \citep{murray98}, or through type I or II orbital 
migration \citep{trilling98, ida04}. These disk interaction scenarios
predict very few high--mass planets ($M_p \gtrsim 2$~\mjup) at small
semi--major axes due to mass--loss suffered from Roche Lobe
overflow. This prediction is in accordance with the observed deficit
of high--mass, short--period planets.

There is another, very different theory for forming short--period
giant planets. In this alternate scenario, planets with initially high
eccentricity and small periastron passages become tidally
circularized at small orbital separations. Using the observed properties
of extrasolar planets, \citet{ford06} 
showed that there exists a sharp cutoff in the mass--period
distribution nearly equal to twice the Roche limit, corresponding
to the ideal circularization distance \citep{faber05, gu03, rasio96}.
If the original eccentricity is induced by 
planet--planet or planet--star interactions
\citep[e.g.][]{weid96,rasio96}, then the expectation is that there
would be fewer high--mass hot jupiters due to their inertial
resistance to scattering, similar to the migration scenario and 
also in agreement with the observed data.  

The Roche limit scales inversely with planet--star mass
ratio and proportionally with the planet radius. Since the planet
radius is nearly constant for $M_P > 1$~\mjup, the tidal
circularization theory helps explain the rarity of massive planets in
extremely short--period orbits. Indeed, the two most recently 
discovered VHJs, HD~189733b and HD~86081b, lie safely beyond the ideal
circularization radius. However, \citet{ford06} caution that the
distance of the 
cutoff was derived under the assumption that the slope follows the Roche
limit since the slope cannot be constrained by the current observational
data. Additional leverage can be gained by increasing the number of
targets monitored in order to find rare high--mass planets close to
the circularization limit, or increasing precision in order to find
low--mass, short--period systems. This highlights the importance of
programs like N2K that are specifically designed to efficiently search
for short--period planets within a large sample of stars.

\acknowledgements 
We would like to thank Ansgar Reiners for his constructive comments
and suggestions on our spectral morphing technique. Thanks to Tim
Robishaw for lending his expertise in data presentation, and for his
many useful IDL Postscript and plotting routines.
We gratefully acknowledge the efforts and dedication
of the Keck Observatory staff. We thank the NOAO and NASA Telescope
assignment committees for generous allocations of telescope time.  
We appreciate funding from NASA grant NNG05GK92G
(to GWM) for supporting this research.   
DAF is a Cottrell Science Scholar of Research Corporation and acknowledges
support from NASA Grant NNG05G164G that made this work possible. 
We also thank NSF for its grants AST-0307493 and AST-9988358 (to
SSV). GWH acknowledges
support from NASA grant NCC5-511 and NSF grant HRD 97-06268.
The authors wish to extend special thanks to those of Hawaiian
ancestry on whose sacred mountain of Mauna Kea we are privileged to be
guests. Without their generous hospitality, the Keck observations
presented herein would not have been possible.

\bibliographystyle{apj}

\begin{thebibliography}{47}
\expandafter\ifx\csname natexlab\endcsname\relax\def\natexlab#1{#1}\fi

\bibitem[{{Ammons} {et~al.}(2006){Ammons}, {Robinson}, {Strader}, {Laughlin},
  {Fischer}, \& {Wolf}}]{ammons06}
{Ammons}, S.~M., {Robinson}, S.~E., {Strader}, J., {Laughlin}, G., {Fischer},
  D., \& {Wolf}, A. 2006, \apj, 638, 1004

\bibitem[{{Bodenheimer} {et~al.}(2003){Bodenheimer}, {Laughlin}, \&
  {Lin}}]{bodenheimer03}
{Bodenheimer}, P., {Laughlin}, G., \& {Lin}, D.~N.~C. 2003, \apj, 592, 555

\bibitem[{{Bouchy} {et~al.}(2004){Bouchy}, {Pont}, {Santos}, {Melo}, {Mayor},
  {Queloz}, \& {Udry}}]{bouchy04}
{Bouchy}, F., {Pont}, F., {Santos}, N.~C., {Melo}, C., {Mayor}, M., {Queloz},
  D., \& {Udry}, S. 2004, \aap, 421, L13

\bibitem[{{Bouchy} {et~al.}(2005){Bouchy}, {Udry}, {Mayor}, {Moutou}, {Pont},
  {Iribarne}, {da Silva}, {Ilovaisky}, {Queloz}, {Santos}, {S{\'e}gransan}, \&
  {Zucker}}]{bouchy05}
{Bouchy}, F., {Udry}, S., {Mayor}, M., {Moutou}, C., {Pont}, F., {Iribarne},
  N., {da Silva}, R., {Ilovaisky}, S., {Queloz}, D., {Santos}, N.~C.,
  {S{\'e}gransan}, D., \& {Zucker}, S. 2005, \aap, 444, L15

\bibitem[{{Butler} {et~al.}(1996){Butler}, {Marcy}, {Williams}, {McCarthy},
  {Dosanjh}, \& {Vogt}}]{butler96}
{Butler}, R.~P., {Marcy}, G.~W., {Williams}, E., {McCarthy}, C., {Dosanjh}, P.,
  \& {Vogt}, S.~S. 1996, \pasp, 108, 500

\bibitem[{{Butler} {et~al.}(2006){Butler}, {Wright}, {Marcy}, {Fischer},
  {Vogt}, {Tinney}, {Jones}, {Carter}, {Johnson}, {McCarthy}, {Munoz}, \&
  {Penny}}]{butler06}
{Butler}, R.~P., {Wright}, J.~T., {Marcy}, G.~W., {Fischer}, D.~A., {Vogt},
  S.~S., {Tinney}, C.~G., {Jones}, H.~R.~A., {Carter}, B.~D., {Johnson}, J.~A.,
  {McCarthy}, C., {Munoz}, M., \& {Penny}, A.~J. 2006, \apj, submitted

\bibitem[{{Eaton} {et~al.}(2003){Eaton}, {Henry}, \& {Fekel}}]{eaton03}
{Eaton}, J.~A., {Henry}, G.~W., \& {Fekel}, F.~C. 2003, {Advantages of
  Automated Observing with Small Telescopes} (The Future of Small Telescopes In
  The New Millennium.~Volume II - The Telescopes We Use), 189--+

\bibitem[{{ESA}(1997)}]{hipp}
{ESA}, . 1997, VizieR Online Data Catalog, 1239, 0

\bibitem[{{Faber} {et~al.}(2005){Faber}, {Rasio}, \& {Willems}}]{faber05}
{Faber}, J.~A., {Rasio}, F.~A., \& {Willems}, B. 2005, Icarus, 175, 248

\bibitem[{{Fischer} \& {Laughlin}(2006)}]{fischer06b}
{Fischer}, A.~D. \& {Laughlin}, G. 2006, in preparation

\bibitem[{{Fischer} {et~al.}(2005){Fischer}, {Laughlin}, {Butler}, {Marcy},
  {Johnson}, {Henry}, {Valenti}, {Vogt}, {Ammons}, {Robinson}, {Spear},
  {Strader}, {Driscoll}, {Fuller}, {Johnson}, {Manrao}, {McCarthy},
  {Mu{\~n}oz}, {Tah}, {Wright}, {Ida}, {Sato}, {Toyota}, \&
  {Minniti}}]{fischer05a}
{Fischer}, D.~A., {Laughlin}, G., {Butler}, P., {Marcy}, G., {Johnson}, J.,
  {Henry}, G., {Valenti}, J., {Vogt}, S., {Ammons}, M., {Robinson}, S.,
  {Spear}, G., {Strader}, J., {Driscoll}, P., {Fuller}, A., {Johnson}, T.,
  {Manrao}, E., {McCarthy}, C., {Mu{\~n}oz}, M., {Tah}, K.~L., {Wright}, J.,
  {Ida}, S., {Sato}, B., {Toyota}, E., \& {Minniti}, D. 2005, \apj, 620, 481

\bibitem[{{Fischer} {et~al.}(2006){Fischer}, {Laughlin}, {Marcy}, {Butler},
  {Vogt}, {Johnson}, {Henry}, {McCarthy}, {Ammons}, {Robinson}, {Strader},
  {Valenti}, {McCullough}, {Charbonneau}, {Haislip}, {Knutson}, {Reichart},
  {McGee}, {Monard}, {Wright}, {Ida}, {Sato}, \& {Minniti}}]{fischer06}
{Fischer}, D.~A., {Laughlin}, G., {Marcy}, G.~W., {Butler}, R.~P., {Vogt},
  S.~S., {Johnson}, J.~A., {Henry}, G.~W., {McCarthy}, C., {Ammons}, M.,
  {Robinson}, S., {Strader}, J., {Valenti}, J.~A., {McCullough}, P.~R.,
  {Charbonneau}, D., {Haislip}, J., {Knutson}, H.~A., {Reichart}, D.~E.,
  {McGee}, P., {Monard}, B., {Wright}, J.~T., {Ida}, S., {Sato}, B., \&
  {Minniti}, D. 2006, \apj, 637, 1094

\bibitem[{{Ford} \& {Rasio}(2006)}]{ford06}
{Ford}, E.~B. \& {Rasio}, F.~A. 2006, \apjl, 638, L45

\bibitem[{{Frink} {et~al.}(2002){Frink}, {Mitchell}, {Quirrenbach}, {Fischer},
  {Marcy}, \& {Butler}}]{frink02}
{Frink}, S., {Mitchell}, D.~S., {Quirrenbach}, A., {Fischer}, D.~A., {Marcy},
  G.~W., \& {Butler}, R.~P. 2002, \apj, 576, 478

\bibitem[{{Frink} {et~al.}(2001){Frink}, {Quirrenbach}, {Fischer}, {R{\"o}ser},
  \& {Schilbach}}]{frink01}
{Frink}, S., {Quirrenbach}, A., {Fischer}, D., {R{\"o}ser}, S., \& {Schilbach},
  E. 2001, \pasp, 113, 173

\bibitem[{{Gaudi} {et~al.}(2005){Gaudi}, {Seager}, \&
  {Mallen-Ornelas}}]{gaudi05}
{Gaudi}, B.~S., {Seager}, S., \& {Mallen-Ornelas}, G. 2005, \apj, 623, 472

\bibitem[{{Gould}(2001)}]{gould01}
{Gould}, A. 2001, \apj, 559, 484

\bibitem[{{Gray}(1992)}]{gray}
{Gray}, D.~F. 1992, The Observation and Analysis of Stellar Photospheres
  (Cambridge University Press), 370--375

\bibitem[{{Gu} {et~al.}(2003){Gu}, {Lin}, \& {Bodenheimer}}]{gu03}
{Gu}, P.-G., {Lin}, D.~N.~C., \& {Bodenheimer}, P.~H. 2003, \apj, 588, 509

\bibitem[{{Henry}(1999)}]{henry99}
{Henry}, G.~W. 1999, \pasp, 111, 845

\bibitem[{{Henry} {et~al.}(2000{\natexlab{a}}){Henry}, {Baliunas}, {Donahue},
  {Fekel}, \& {Soon}}]{henry00a}
{Henry}, G.~W., {Baliunas}, S.~L., {Donahue}, R.~A., {Fekel}, F.~C., \& {Soon},
  W. 2000{\natexlab{a}}, \apj, 531, 415

\bibitem[{{Henry} {et~al.}(2000{\natexlab{b}}){Henry}, {Marcy}, {Butler}, \&
  {Vogt}}]{henry00b}
{Henry}, G.~W., {Marcy}, G.~W., {Butler}, R.~P., \& {Vogt}, S.~S.
  2000{\natexlab{b}}, \apjl, 529, L41

\bibitem[{{Henry} {et~al.}(1995){Henry}, {Fekel}, \& {Hall}}]{henry95}
{Henry}, G.~W., {Fekel}, F.~C., \& {Hall}, D.~S. 1995, \aj, 110, 2926

\bibitem[{{Ida} \& {Lin}(2004)}]{ida04}
{Ida}, S. \& {Lin}, D.~N.~C. 2004, \apj, 604, 388

\bibitem[{{Jansson}(1995)}]{jansson}
{Jansson}, P. 1995, Deconvolution: With Applications in Spectroscopy (Academic
  Press)

\bibitem[{{Konacki} {et~al.}(2003){Konacki}, {Torres}, {Jha}, \&
  {Sasselov}}]{konacki03}
{Konacki}, M., {Torres}, G., {Jha}, S., \& {Sasselov}, D.~D. 2003, \nat, 421,
  507

\bibitem[{{Marcy} {et~al.}(2005{\natexlab{a}}){Marcy}, {Butler}, {Fischer},
  {Vogt}, {Wright}, {Tinney}, \& {Jones}}]{marcy05a}
{Marcy}, G., {Butler}, R.~P., {Fischer}, D., {Vogt}, S., {Wright}, J.~T.,
  {Tinney}, C.~G., \& {Jones}, H.~R.~A. 2005{\natexlab{a}}, Progress of
  Theoretical Physics Supplement, 158, 24

\bibitem[{{Marcy} {et~al.}(2005{\natexlab{b}}){Marcy}, {Butler}, {Vogt},
  {Fischer}, {Henry}, {Laughlin}, {Wright}, \& {Johnson}}]{marcy05b}
{Marcy}, G.~W., {Butler}, R.~P., {Vogt}, S.~S., {Fischer}, D.~A., {Henry},
  G.~W., {Laughlin}, G., {Wright}, J.~T., \& {Johnson}, J.~A.
  2005{\natexlab{b}}, \apj, 619, 570

\bibitem[{{Murray} {et~al.}(1998){Murray}, {Hansen}, {Holman}, \&
  {Tremaine}}]{murray98}
{Murray}, N., {Hansen}, B., {Holman}, M., \& {Tremaine}, S. 1998, Science, 279,
  69

\bibitem[{{Noyes} {et~al.}(1984){Noyes}, {Hartmann}, {Baliunas}, {Duncan}, \&
  {Vaughan}}]{noyes84}
{Noyes}, R.~W., {Hartmann}, L.~W., {Baliunas}, S.~L., {Duncan}, D.~K., \&
  {Vaughan}, A.~H. 1984, \apj, 279, 763

\bibitem[{{Paulson} {et~al.}(2004){Paulson}, {Saar}, {Cochran}, \&
  {Henry}}]{paulson04}
{Paulson}, D.~B., {Saar}, S.~H., {Cochran}, W.~D., \& {Henry}, G.~W. 2004, \aj,
  127, 1644

\bibitem[{{Queloz} {et~al.}(2001){Queloz}, {Henry}, {Sivan}, {Baliunas},
  {Beuzit}, {Donahue}, {Mayor}, {Naef}, {Perrier}, \& {Udry}}]{queloz01}
{Queloz}, D., {Henry}, G.~W., {Sivan}, J.~P., {Baliunas}, S.~L., {Beuzit},
  J.~L., {Donahue}, R.~A., {Mayor}, M., {Naef}, D., {Perrier}, C., \& {Udry},
  S. 2001, \aap, 379, 279

\bibitem[{{Rasio} \& {Ford}(1996)}]{rasio96}
{Rasio}, F.~A. \& {Ford}, E.~B. 1996, Science, 274, 954

\bibitem[{{Robinson} {et~al.}(2006){Robinson}, {Strader}, {Ammons}, {Laughlin},
  \& {Fischer}}]{robinson06}
{Robinson}, S.~E., {Strader}, J., {Ammons}, S.~M., {Laughlin}, G., \&
  {Fischer}, D. 2006, \apj, 637, 1102

\bibitem[{{Sato} {et~al.}(2005{\natexlab{a}}){Sato}, {Fischer}, {Henry},
  {Laughlin}, {Butler}, {Marcy}, {Vogt}, {Bodenheimer}, {Ida}, {Toyota},
  {Wolf}, {Valenti}, {Boyd}, {Johnson}, {Wright}, {Ammons}, {Robinson},
  {Strader}, {McCarthy}, {Tah}, \& {Minniti}}]{sato05}
{Sato}, B., {Fischer}, D.~A., {Henry}, G.~W., {Laughlin}, G., {Butler}, R.~P.,
  {Marcy}, G.~W., {Vogt}, S.~S., {Bodenheimer}, P., {Ida}, S., {Toyota}, E.,
  {Wolf}, A., {Valenti}, J.~A., {Boyd}, L.~J., {Johnson}, J.~A., {Wright},
  J.~T., {Ammons}, M., {Robinson}, S., {Strader}, J., {McCarthy}, C., {Tah},
  K.~L., \& {Minniti}, D. 2005{\natexlab{a}}, \apj, 633, 465

\bibitem[{{Sato} {et~al.}(2005{\natexlab{b}}){Sato}, {Kambe}, {Takeda},
  {Izumiura}, {Masuda}, \& {Ando}}]{sato05b}
{Sato}, B., {Kambe}, E., {Takeda}, Y., {Izumiura}, H., {Masuda}, S., \& {Ando},
  H. 2005{\natexlab{b}}, \pasj, 57, 97

\bibitem[{{Seagroves} {et~al.}(2003){Seagroves}, {Harker}, {Laughlin}, {Lacy},
  \& {Castellano}}]{seagroves03}
{Seagroves}, S., {Harker}, J., {Laughlin}, G., {Lacy}, J., \& {Castellano}, T.
  2003, \pasp, 115, 1355

\bibitem[{{Starck} {et~al.}(2002){Starck}, {Pantin}, \& {Murtagh}}]{starck02}
{Starck}, J.~L., {Pantin}, E., \& {Murtagh}, F. 2002, \pasp, 114, 1051

\bibitem[{{Trilling} {et~al.}(1998){Trilling}, {Benz}, {Guillot}, {Lunine},
  {Hubbard}, \& {Burrows}}]{trilling98}
{Trilling}, D.~E., {Benz}, W., {Guillot}, T., {Lunine}, J.~I., {Hubbard},
  W.~B., \& {Burrows}, A. 1998, \apj, 500, 428

\bibitem[{{Udry} {et~al.}(2003){Udry}, {Mayor}, {Clausen}, {Freyhammer},
  {Helt}, {Lovis}, {Naef}, {Olsen}, {Pepe}, {Queloz}, \& {Santos}}]{udry03}
{Udry}, S., {Mayor}, M., {Clausen}, J.~V., {Freyhammer}, L.~M., {Helt}, B.~E.,
  {Lovis}, C., {Naef}, D., {Olsen}, E.~H., {Pepe}, F., {Queloz}, D., \&
  {Santos}, N.~C. 2003, \aap, 407, 679

\bibitem[{{Valenti} \& {Fischer}(2005)}]{valenti05}
{Valenti}, J.~A. \& {Fischer}, D.~A. 2005, \apjs, 159, 141

\bibitem[{{Vogt} {et~al.}(1994){Vogt}, {Allen}, {Bigelow}, {Bresee}, {Brown},
  {Cantrall}, {Conrad}, {Couture}, {Delaney}, {Epps}, {Hilyard}, {Hilyard},
  {Horn}, {Jern}, {Kanto}, {Keane}, {Kibrick}, {Lewis}, {Osborne},
  {Pardeilhan}, {Pfister}, {Ricketts}, {Robinson}, {Stover}, {Tucker}, {Ward},
  \& {Wei}}]{vogt94}
{Vogt}, S.~S., {Allen}, S.~L., {Bigelow}, B.~C., {Bresee}, L., {Brown}, B.,
  {Cantrall}, T., {Conrad}, A., {Couture}, M., {Delaney}, C., {Epps}, H.~W.,
  {Hilyard}, D., {Hilyard}, D.~F., {Horn}, E., {Jern}, N., {Kanto}, D.,
  {Keane}, M.~J., {Kibrick}, R.~I., {Lewis}, J.~W., {Osborne}, J.,
  {Pardeilhan}, G.~H., {Pfister}, T., {Ricketts}, T., {Robinson}, L.~B.,
  {Stover}, R.~J., {Tucker}, D., {Ward}, J., \& {Wei}, M.~Z. 1994, in Proc.
  SPIE Instrumentation in Astronomy VIII, David L. Crawford; Eric R. Craine;
  Eds., Volume 2198, p. 362, ed. D.~L. {Crawford} \& E.~R. {Craine}, 362--+

\bibitem[{{Vogt} {et~al.}(2000){Vogt}, {Marcy}, {Butler}, \& {Apps}}]{vogt00}
{Vogt}, S.~S., {Marcy}, G.~W., {Butler}, R.~P., \& {Apps}, K. 2000, \apj, 536,
  902

\bibitem[{{Weidenschilling} \& {Marzari}(1996)}]{weid96}
{Weidenschilling}, S.~J. \& {Marzari}, F. 1996, \nat, 384, 619

\bibitem[{{Winn} {et~al.}(2005){Winn}, {Noyes}, {Holman}, {Charbonneau},
  {Ohta}, {Taruya}, {Suto}, {Narita}, {Turner}, {Johnson}, {Marcy}, {Butler},
  \& {Vogt}}]{winn05}
{Winn}, J.~N., {Noyes}, R.~W., {Holman}, M.~J., {Charbonneau}, D., {Ohta}, Y.,
  {Taruya}, A., {Suto}, Y., {Narita}, N., {Turner}, E.~L., {Johnson}, J.~A.,
  {Marcy}, G.~W., {Butler}, R.~P., \& {Vogt}, S.~S. 2005, \apj, 631, 1215

\bibitem[{{Wright}(2005)}]{wright05}
{Wright}, J.~T. 2005, \pasp, 117, 657

\bibitem[{{Yi} {et~al.}(2001){Yi}, {Demarque}, {Kim}, {Lee}, {Ree}, {Lejeune},
  \& {Barnes}}]{yi02}
{Yi}, S., {Demarque}, P., {Kim}, Y.-C., {Lee}, Y.-W., {Ree}, C.~H., {Lejeune},
  T., \& {Barnes}, S. 2001, \apjs, 136, 417

\end{thebibliography}

\clearpage
\begin{deluxetable}{lllll}
\tablecaption{Stellar Parameters\label{stars}}
\tablewidth{0pt}
\tablehead{\colhead{Parameter}  & 
\colhead{HD 86081} & 
\colhead{HD 224693} &
\colhead{HD 33283} \\
}
\startdata
V              & \vmaga         & \vmagc      & \vmagd  \\
$M_V$          & \mva           & \mvc        & \mvd    \\
B-V            & \bva           & \bvc        & \bvd    \\
Spectral Type  & \spa           & \spc        & \spd    \\
Distance (pc)  & \da            & \dc         & \dd     \\
${\rm [Fe/H]}$ & \fea~(0.03)    & \fec~(0.03) & \fed~(0.03)\\
$T_{eff}$~(K)  & \teffa~(44)    & \teffc~(44) & \teffd~(44)  \\
\vsini~\ks     & \va~(0.5)      & \vc~(0.5)   & \vd~(0.5)  \\
\logg          & \ga~(0.07)     & \gc~(0.07)  & \gd~(0.07) \\
$M_{*}$~(\msun) & \msa~(0.05)    & \msc~(0.1)  & \msd~(0.1) \\
$R_{*}$~(\rsun) & \rsa~(0.1)    & \rsc~(0.3)  & \rsd~(0.1) \\
$S_{HK}$       & \shka          & \shkc       & \shkd      \\
$\log R'_{HK}$ & \rhka          & \rhkc       & \rhkd      \\
$P_{rot}$~(d)  & \pra           & \prc        &  \prd \\
\enddata
\end{deluxetable}

\clearpage
\begin{deluxetable}{lrc}
\tablecaption{Radial Velocities for HD~86081\label{vel86081}}
\tablewidth{0pt}
\tablehead{
\colhead{JD} &
\colhead{RV} &
\colhead{Uncertainty} \\
\colhead{-2440000} &
\colhead{(m~s$^{-1}$)} &
\colhead{(m~s$^{-1}$)} 
}
\startdata
13694.156 & -168.30 & 4.20 \\
13695.155 &  129.92 & 4.29 \\
13697.155 &   52.46 & 4.77 \\
13746.991 &  151.71 & 3.40 \\
13747.997 & -184.74 & 3.20 \\
13748.936 &  201.56 & 2.72 \\
13749.891 & -209.26 & 3.06 \\
13750.907 &  189.65 & 2.12 \\
13750.936 &  195.23 & 1.85 \\
13751.883 & -170.66 & 2.09 \\
13751.937 & -196.19 & 3.68 \\
13752.948 &  165.28 & 2.30 \\
13752.986 &  172.90 & 2.22 \\
13753.060 &  197.83 & 2.11 \\
13753.124 &  201.11 & 2.25 \\
13753.912 & -121.22 & 2.83 \\
13753.934 & -132.36 & 2.27 \\
13753.962 & -142.86 & 1.97 \\
13754.033 & -171.86 & 2.17 \\
13754.069 & -185.19 & 2.07 \\
13775.897 & -142.57 & 3.60 \\
13776.835 &  180.54 & 3.50 \\
13777.920 & -191.64 & 3.46 \\
13778.927 &  193.64 & 3.43 \\
13779.925 & -207.33 & 3.80 \\
13781.064 &  190.15 & 3.64 \\
\enddata
\end{deluxetable}

\begin{deluxetable}{lrc}
\tablecaption{Radial Velocities for HD~224693\label{vel224693}}
\tablewidth{0pt}
\tablehead{
\colhead{JD} &
\colhead{RV} &
\colhead{Uncertainty} \\
\colhead{-2440000} &
\colhead{(m~s$^{-1}$)} &
\colhead{(m~s$^{-1}$)} 
}
\startdata
13191.097 &   47.16 & 5.09 \\
13198.131 &   20.71 & 5.54 \\
13199.091 &   17.74 & 5.50 \\
13200.042 &    9.35 & 5.06 \\
13367.705 &  -37.79 & 2.65 \\
13368.713 &  -26.67 & 3.26 \\
13369.714 &  -26.54 & 2.50 \\
13370.736 &  -13.68 & 11.3 \\
13550.112 &  -20.60 & 3.48 \\
13551.122 &  -18.08 & 3.25 \\
13552.090 &  -27.37 & 3.38 \\
13571.088 &   32.34 & 3.31 \\
13603.089 &  -15.32 & 3.06 \\
13604.062 &  -21.39 & 2.87 \\
13605.021 &  -24.05 & 2.87 \\
13692.906 &   -6.92 & 1.97 \\
13693.850 &   -3.39 & 1.95 \\
13694.762 &    3.36 & 2.11 \\
13695.757 &    9.33 & 2.09 \\
13696.751 &   23.32 & 2.01 \\
13723.810 &   31.15 & 2.65 \\
13724.756 &   35.05 & 2.68 \\
13746.696 &   -6.67 & 3.14 \\
13752.761 &   40.52 & 2.67 \\
\enddata
\end{deluxetable}

\clearpage
\begin{deluxetable}{lrc}
\tablecaption{Radial Velocities for HD~33283\label{vel33283}}
\tablewidth{0pt}
\tablehead{
\colhead{JD} &
\colhead{RV} &
\colhead{Uncertainty} \\
\colhead{-2440000} &
\colhead{(m~s$^{-1}$)} &
\colhead{(m~s$^{-1}$)} 
}
\startdata
13014.852 &   10.57 & 5.45 \\
13015.858 &    1.88 & 5.35 \\
13016.850 &  -11.57 & 4.98 \\
13071.767 &  -29.19 & 4.85 \\
13072.843 &  -31.26 & 5.07 \\
13368.955 &    4.10 & 4.94 \\
13369.808 &    4.87 & 4.99 \\
13397.796 &   -8.86 & 4.52 \\
13398.870 &  -20.06 & 5.14 \\
13428.792 &   17.14 & 4.50 \\
13605.139 &   10.66 & 4.35 \\
13692.917 &  -12.34 & 3.93 \\
13693.979 &   -2.42 & 4.23 \\
13694.975 &   -0.03 & 4.35 \\
13695.994 &    6.55 & 4.33 \\
13696.981 &    9.04 & 4.26 \\
13723.925 &    9.44 & 4.49 \\
13724.932 &    4.19 & 4.53 \\
13746.852 &  -15.10 & 4.81 \\
13747.891 &   -2.94 & 4.60 \\
13748.809 &   -0.13 & 4.37 \\
13749.824 &    5.66 & 4.27 \\
13750.822 &   12.91 & 4.14 \\
13751.875 &   14.61 & 3.34 \\
13752.878 &   13.17 & 3.89 \\
\enddata
\end{deluxetable}

\begin{deluxetable}{lll}
\tablecaption{Orbital Parameters for HD~86081 From Observed and
  Morphed Templates\label{param86081}}
\tablewidth{0pt}
\tablehead{
\colhead{Parameter} &
\colhead{Observed} &
\colhead{Morphed} \\
}
\startdata
P~(d)               & \pa~(\pea)     & \pam~(\peam)     \\
T$_p$\tablenotemark{a}~(JD)       & \tpa~(\tpea) & \tpam~(\tpeam) \\
e                   & \ea~(\eea)     & \eam~(\eeam)     \\
K$_1$~(\ms)         & \ka~(\kea)     & \kam~(\keam)     \\
$\omega$~(deg)      & \oma~(\omea)   & \omam~(\omeam)   \\
\msini~($M_{Jup}$)  & \ma           & \mam          \\
\asini~(AU)         & \aa           & \aam          \\
Fit RMS~(\ms)       & \rmsa          & \rmsam        \\
$\sqrt{\chi_\nu^2}$ & \chia   & \chiam       \\
N$_{obs}$           & \nobsa             & \nobsam   \\
\enddata
\tablenotetext{a}{Time of periastron passage.}
\end{deluxetable}

\clearpage
\begin{deluxetable}{lll}
\tablecaption{Orbital Parameters for HD~224693 and HD~33283\label{paramothers}}
\tablewidth{0pt}
\tablehead{
\colhead{Parameter} &
\colhead{HD~224693} &
\colhead{HD~33283}  \\
}
\startdata
P~(d)               & \pc~(\pec)     & \pd~(\ped)     \\
T$_p$\tablenotemark{a}~(JD)       & \tpc~(\tpec) & \tpd~(\tped) \\
e                   & \ec~(\eec)     & \ed~(\eed)     \\
K$_1$~(\ms)         & \kc~(\kec)     & \kd~(\ked)     \\
$\omega$~(deg)      & \omc~(\omec)   & \omd~(\omed)   \\
\msini~($M_{Jup}$)   & \mc           & \md        \\
\asini~(AU)         & \ac            & \ad          \\
Fit RMS~(\ms)       & \rmsc          & \rmsd       \\
$\sqrt{\chi_\nu^2}$ & \chic          & \chid       \\
N$_{obs}$           & \nobsc         & \nobsd      \\
\enddata
\tablenotetext{a}{Time of periastron passage.}
\end{deluxetable}

\clearpage
\begin{deluxetable}{ccc}
\tablewidth{0pt}
\tablecaption{Photometric Observations of HD~86081\label{photmeas}}
\tablehead{
\colhead{Observation Date} & \colhead{$\Delta (b+y)/2$} \\
\colhead{(HJD $-$ 2,400,000)} & \colhead{(mag)}
}
\startdata
53,739.9097 & 1.2972 \\
53,739.9168 & 1.2987 \\
53,739.9654 & 1.2985 \\
53,740.8213 & 1.2976 \\
53,740.9088 & 1.2964 \\
\enddata
\tablecomments{Table \ref{photmeas}A, along with Table
  \ref{photmeas}B (for HD~224693), is presented in its entirety in the
  electronic edition of the Astrophysical Journal.  A portion is shown
  here for guidance regarding its form and content.}
\end{deluxetable}

\clearpage
\begin{deluxetable}{ccccccc}
\tabletypesize{\small}
\tablewidth{0pt}
\tablecaption{Photometric Results for the Planetary Host Stars \label{photres}}
\tablehead{
\colhead{} & \colhead{Date Range} & \colhead{} & \colhead{$\sigma$} & \colhead{Semi-Amplitude} & \colhead{Transit Probability} & \colhead{} \\
\colhead{Star} & \colhead{(HJD $-$ 2,450,000)} & \colhead{$N_{obs}$} & \colhead{(mag)} & \colhead{(mag)} & \colhead{(\%)} & \colhead{Transits} \\
\colhead{(1)} & \colhead{(2)} & \colhead{(3)} & \colhead{(4)} & \colhead{(5)} & \colhead{(6)} & \colhead{(7)}
}
\startdata
 HD 86081 & 3739--3777 & 135 & 0.0019 & 0.0004 $\pm$ 0.0002 & 17.4 & no \\
HD 224693 & 3691--3733 &  22 & 0.0018 & 0.0003 $\pm$ 0.0004 &  4.2 &  ? \\
\enddata
\end{deluxetable}
\end{document}